\documentclass[]{aa}  

\usepackage{graphicx}
\usepackage{txfonts}
\usepackage{color}
\usepackage{comment}
\usepackage[colorlinks=true, allcolors=blue]{hyperref}

\newcommand{\kms}{\ensuremath{\mathrm{km}\,\mathrm{s}^{-1}}}

\newcommand{\hi } {{\rm H}\,{\small\rm I}}

\newcommand{\cii } {[{\rm C}\,{\small\rm II}]}
\newcommand{\nii } {[{\rm N}\,{\small\rm II}]}

\newcommand{\oiii } {[{\rm O}\,{\small\rm III}]}
\newcommand{\bb } {\textsc{$^{\rm 3D}$Barolo}}

\begin{document} 

   \title{TRICEPS. I. Tracing cold gas, dust, and stars at sub-kpc scales in massive star-forming galaxies at $z\simeq4-5$}

   \titlerunning{TRICEPS. I. Tracing cold gas, dust, and stars at sub-kpc scales in massive star-forming galaxies at $z\simeq4-5$}

   \author{Federico Lelli\inst{1,2}
        \and
        Lingrui Lin\inst{2,3,4}
        \and
        Zhi-Yu Zhang\inst{3,4}
        \and
        Cecilia Bacchini\inst{5, 6}
        \and
        Davide Bevacqua\inst{7}
        \and
        Joe Bhangal\inst{8}
        \and
        Marco Castellano\inst{7}
        \and        
        Carlos De Breuck\inst{9}
        \and
        Giovanni Gandolfi\inst{7}
        \and
        Allison W. S. Man\inst{8}
        \and
        Antonino Marasco\inst{6}
        \and
        Diego Paris\inst{7}
        \and
        Borja P\'erez-D\'iaz\inst{7}
        \and
        Paola Santini\inst{7}
}
\institute{Dipartimento di Fisica, Università degli Studi di Cagliari, Monserrato 09042, Italy; \email{federico.lelli@dsf.inaf.it}
    \and
    INAF $-$ Arcetri Astrophysical Observatory, Largo E. Fermi 5, 50125, Florence, Italy;
    \and
    School of Astronomy and Space Science, Nanjing University, Nanjing 210023, P.R. China
    \and
    Key Laboratory of Modern Astronomy and Astrophysics, Nanjing University, Ministry of Education, Nanjing 210023, P.R. China
    \and
    DARK, Niels Bohr Institute, University of Copenhagen, Jagtvej 155, 2200, Copenhagen, Denmark
    \and
    INAF $-$ Padova Astronomical Observatory, Vicolo dell’Osservatorio 5, 35122 Padova, Italy
    \and
    INAF $-$ Rome Astronomical Observatory, Via Frascati 33, 00078, Rome, Italy
    \and
    Department of Physics and Astronomy, University of British Columbia, 6224 Agricultural Road, Vancouver BC, V6T 1Z1, Canada
    \and
    European Southern Observatory, Karl-Schwarzschild-Strasse 2, D-85748 Garching bei München, Germany
    } 
   \date{Received XX/XX/XXXX; accepted XX/XX/XXXX}
 
  \abstract
{We introduce the TRICEPS (Tracing Resolved Ionized Carbon Emission in Primeval Systems) survey. TRICEPS focuses on a sample of 16 massive galaxies at $z=4-5$ with top-quality data: (i) ALMA band-7 observations of the \cii-158 $\mu$m line, tracing the cold gas distribution and kinematics at spatial resolutions of about $0.1''$ ($\sim0.8$ kpc), (ii) ALMA maps of the rest-frame 160 $\mu$m continuum emission, tracing the dust distribution at similar spatial resolutions, (iii) multi-band JWST images from NIRCam and MIRI, tracing the stellar mass distribution and star formation activity at spatial resolutions of $0.1''-0.2''$, and (iv) multi-line CO observations from ALMA and the VLA, including the CO(1-0) line, tracing the molecular gas content. In this first paper, we describe the TRICEPS scientific goals, the basic properties of the galaxy sample, and the ALMA band-7 observations. TRICEPS galaxies were selected based only on bright \cii\ fluxes (>2 Jy \kms). They lie on the top-end of the star-forming main sequence at $z=4-5$, having stellar masses between about $10^{10.5}-10^{12}$ M$_\odot$ and star-formation rates between about $100-3000$ M$_\odot$ yr$^{-1}$. To date, TRICEPS is the largest sample of galaxies at $z=4-5$ for which the gas distribution and kinematics are exceptionally well resolved: on average, 46 resolution elements per galaxy with signal-to-noise ratio higher than 3. By survey design, half of the TRICEPS targets are in galaxy pairs; clear signs of tidal interactions are present in three of the four pairs. In all galaxies, the \cii\ velocity fields display regular rotation, albeit the interacting galaxies tend to be somewhat more disturbed. The TRICEPS observations suggest that regular rotation is ubiquitous among massive star-forming galaxies when the Universe was less than $\sim$1.5 Gyr old.}


   \maketitle

\section{Introduction}\label{sec:intro}

During the past decade, the Atacama Large Millimeter Array (ALMA) has opened a new window to study cold gas in galaxies at $z\gtrsim4$ thanks to the \cii\ emission line at 158 $\mu$m \citep{Swinbank2012, Wagg2012, Carilli2013, Carniani2013, Riechers2013, DeBreuck2014}. The \cii\ line is the fine-structure  $^{2}P^{0}_{3/2}\rightarrow^{2}P^{0}_{1/2}$ transition of singly ionized carbon, C$^{+}$. It is one of the brightest emission lines in the far infrared (FIR) and a major cooling channel of the interstellar medium (ISM). It is thought to be mostly produced in photodissociation regions \citep{Crawford1985, Stacey1991} with a possible contribution from X-ray dominated regions in active galactic nuclei \citep[AGN,][]{Stacey2010}. Due to the low ionization potential of neutral carbon of 11.26 eV (lower than that of atomic hydrogen of 13.6 eV), C$^{+}$ can be present through the different phases of the ISM: molecular, atomic, and ionized. Indeed, collisions with molecular hydrogen (H$_2$), atomic hydrogen (\hi), and free electrons can excite C$^{+}$ gas and produce \cii\ emission, which is therefore a multiphase gas tracer.

Due to its multi-tracer nature, the \cii\ luminosity ($L_{\cii}$) has been used to measure various physical quantities: star-formation rates \citep[SFRs,][]{DeLooze2014, HerreraCamus2015}, H$_2$ masses \citep{Zanella2018}, and \hi\ masses \citep{Heintz2021, Heintz2022}. In the nearby Universe ($z\simeq0$), in-depth studies of spiral galaxies using multiple emission lines (e.g., \hi, CO, \nii, and \cii) found that the \cii\ emission is mostly associated with atomic and molecular hydrogen, whereas the association with ionized hydrogen is almost negligible (see \citealt{Pineda2013} for the Milky Way, \citealt{Kapala2015} for M31, and \citealt{Tarantino2021} for M101 and NGC\,6946). If this holds for the general galaxy population, the $L_{\cii}-\rm{SFR}$ relation would then be a consequence of the global Kennicutt-Schmidt relation between the cold gas mass $M_{\rm gas}$ (atomic plus molecular) and the SFR \citep{Kennicutt1998}, corroborating that the \cii\ emission is effectively tracing the fuel for star formation: neutral gas cooling down. Although the relative contributions of atomic, molecular, and ionized gas to the \cii\ emission may be different at high $z$ with respect to low $z$, the \cii\ line is clearly a good tracer of cold gas with temperature $T\lesssim10^4$ K \citep{CarilliWalter2013}.

During the past decade, several ALMA surveys studied the \cii \ emission in galaxies at $z>4$. Broadly speaking, three main observing strategies have been used: (i) blind low-resolution surveys in well-studied extragalactic fields, such as the large program ASPECS \citep{Walter2016}; (ii) targeted surveys at moderate angular resolutions ($0.5''-1''$) of significant samples ($\sim$50$-$100 galaxies) such as the large programs ALPINE at $4<z<6$ \citep{LeFevre2020} and REBELS at $z>6.5$ \citep{Bouwens2022}; and (iii) targeted surveys at high angular resolutions ($0.1''-0.2''$) of individual unlensed galaxies \citep{Tadaki2018, Neeleman2020, Lelli2021, Tsuki2021} or small samples (4-5 objects) of lensed galaxies \citep{Rizzo2020, Rizzo2021}. An intermediate approach in-between (ii) and (iii) has been adopted by the ALMA large program CRISTAL \citep{HerreraCamus2025A&A...699A..80H}, which obtained \cii\ data with a mean resolution of $\sim0.3''$ for a sample of 19 galaxies at $4<z<6$. In this paper, we present \cii\ data with a mean resolution of $\sim0.1''$ for a sample of 16 massive galaxies at $4<z<5$; similarities and differences with respect to the CRISTAL survey are discussed in detail in Sect.\,\ref{sec:sampleprop}.

In general, these different approaches are highly complementary: while large samples with low-resolution \cii\ data provide a global view of the gas content in the overall galaxy population, small samples with high-resolution data allow us to spatially resolve the gas distribution and kinematics. Resolving the \cii\ emission at sub-kpc spatial scales is necessary to study the overall gas morphology (e.g., clumpy versus smooth distributions), trace well-sampled \cii\ surface brightness profiles, and investigate the physical mechanisms responsible for the \cii\ emission. In addition, high-resolution \cii\ data are needed to unambiguously distinguish rotating disks from more complex kinematic configurations, such as mergers, gas outflows, and/or gas inflows, which is a very difficult task with low-resolution data \citep[e.g.,][]{Jones2021, Rizzo2022}.

Here we introduce the TRICEPS (Tracing Resolved Ionized Carbon Emission in Primeval Systems) survey, which provides high-quality, multi-wavelength, multi-facility observations for a sample of 16 massive galaxies at $4<z<5$, when the age of the Universe was between about 1.2 and 1.5 Gyr. In particular, we obtained the following observations:
\begin{itemize}
    \item ALMA band 7 data of the \cii\ line at a mean angular resolutions of $0.1''$, corresponding to about 0.8 kpc, resolving the gas distribution and kinematics with plenty of resolution elements (program ID: 2019.1.01587.S, PI: F. Lelli). The same observations also provide continuum maps of the rest-frame 160 $\mu$m emission, which traces dust heated by young stars.
    \item Multi-band imaging with the \emph{James Webb Space Telescope} (JWST, program ID 3954, PI: F. Lelli). These images probe the rest-frame UV-to-NIR emission at similar angular resolutions as the ALMA data, allowing us to measure the spatial distribution of SFRs and stellar mass ($M_\star$) from fitting the spectral energy distribution (SED).
    \item CO($1-0$) data using the \emph{Jansky Very Large Array} (VLA, program ID 24A-296, PI: F. Lelli) and CO data of higher $J$ transitions using various ALMA bands (program ID 2025.1.00827.S, PI: F. Lelli). These observations are in progress and will allow building the CO spectral line energy distribution and accurately measure the molecular gas mass.
\end{itemize}
The TRICEPS survey will therefore provide a panchromatic view of gas, dust, and stars in a significant sample of massive galaxies at $z\simeq4-5$. In this first paper of the TRICEPS series, we present the selection of the galaxy sample and its basic properties (Sect.\,\ref{sec:sample}), describe the new ALMA band 7 observations and data reduction (Sect.\,\ref{sec:ALMA}), and discuss basic results on the \cii\ morphology and kinematics (Sect.\,\ref{sec:results}). Additional observations and analyses are presented in the following companion papers:
\begin{itemize}
    \item In Marasco et al. (submitted, hereafter Paper II), we present the new JWST observations, their photometric analysis, and the measurements of total $M_\star$ and SFRs from SED fitting.
    \item In Lin et al. (submitted, hereafter Paper III), we present the kinematic analysis of the \cii\ data, finding that all TRICEPS galaxies possess a rotation-supported gas disk. The analysis provides the disk geometry (center, position angle, inclination angle), rotation curves, and velocity dispersions.
    \item In Bhangal et al. (submitted, hereafter Paper IV), we use the JWST images to study the environment of TRICEPS galaxies, finding that at least eleven of them reside in significant overdensities (proto-clusters).
    \item In Lu et al. (submitted, hereafter Paper V), we use the ALMA data to study the spatially resolved Kennicutt-Schmidt relation \citep{Kennicutt1998}, linking the gas surface density with the SFR surface density.
\end{itemize}
Additional papers on various science goals are in preparation, as we anticipate in Section\,\ref{sec:prospects} of this paper.

Throughout all the TRICEPS papers, we adopt a flat $\Lambda$ cold dark matter ($\Lambda$CDM) cosmology with $H_{0}= 67.4$ km s$^{-1}$ Mpc$^{-1}$, $\Omega_{\rm m} = 0.315$, and $\Omega_{\Lambda} = 0.685$ \citep{Planck2018}. In this cosmology, at a reference $z=4.5$, the age of the Universe is 1.3 Gyr, the light travel time is 12.5 Gyr, and an angular resolution of $0.1''$ corresponds to 675 pc.

\section{The TRICEPS sample}\label{sec:sample}

\subsection{Sample selection}

The TRICEPS project was designed in 2019 and accepted for ALMA Cycle 7 (project ID: 2019.1.01587.S, PI: F. Lelli). The starting idea was to obtain follow-up observations at high spatial resolutions for galaxies at $z\simeq4-5$ with known \cii\ fluxes. Next, the project expanded by obtaining JWST imaging and multi-line CO observations for the same galaxies.

The sample selection was defined in spring 2019 before large low-resolution \cii\ samples, such as ALPINE, were publicly available. We searched the ALMA archive for \cii\ detections at $z=4-5$ with published fluxes and redshifts in peer-reviewed publications. The low-redshift edge ($z>4$) is motivated by the difficulty of obtaining high-frequency observations, which require outstanding weather conditions at the ALMA site. The high-redshift edge ($z<5$), instead, is motivated by the need of minimizing cosmological surface brightness dimming in spatially resolved observations. We also excluded gravitationally lensed galaxies because the interpretation of the \cii\ kinematics is more complex due to, e..g, the degeneracy with the lens parameters \citep{Rizzo2018}. Finally, we selected galaxies with total \cii\ fluxes ($I_{\cii}$) above 2 Jy\,km\,s$^{-1}$ to have sensible integration times with ALMA (see Sect.\,\ref{sec:newobs}). At $z=4.5$, the latter criterion corresponds to $L_{\cii}\gtrsim 1.3 \times 10^9$ L$_\odot$, which in turn corresponds to $M_{\rm gas} \gtrsim 4 \times 10^{10}$ M$_\odot$ for a fiducial $\alpha_{\cii}$ conversion factor of 30 M$_\odot$/L$_\odot$ \citep{Zanella2018}. Bright galaxies are the obvious starting point for a high-resolution \cii\ survey that aims to obtain the best-possible data to resolve the gas distribution and kinematics with plenty of independent elements.

\begin{table*}
\caption[]{General properties of galaxies in the TRICEPS sample.}
\centering
\footnotesize
\label{tab:sample}
\renewcommand{\arraystretch}{1.3}
\resizebox{\textwidth}{!}{%
\begin{tabular}{lcccccccccc}
\hline
Galaxy & Type & $z$ & R.A.& Dec. & $I_{\cii}$ & $L_{\cii}$ & $I_{\rm 160 \mu m}$ & $L_{\rm 160 \mu m}$ & $\log(M_\star)$ & $\log(\rm{SFR})$ \\
       &      &     & deg & deg  & Jy \kms & $10^{9}$ L$_{\odot}$ & mJy & L$_{\odot}$ Hz$^{-1}$ & M$_{\odot}$ & M$_{\odot}$ yr$^{-1}$ \\
\hline
  ALESS-73.1 & SMG & 4.755 &  53.12207 & -27.93878 & $7.6\pm0.4$   & $5.2\pm0.2$   &  $7.7\pm0.1$  & $0.84\pm0.02$ & $10.70^{+0.12}_{-0.17}$ & $2.19^{+0.55}_{-0.18}$ \\
     AzTEC-1 & SMG & 4.342 & 149.92856 &  +2.49395 & $18.8\pm0.8$  & $11.4\pm0.5$  &  $19.5\pm0.4$   & $1.86\pm0.04$ & $10.83^{+0.32}_{-0.20}$ & $2.45^{+0.20}_{-0.17}$ \\
   AzTEC-159 & SMG & 4.567 & 149.87669 &  +1.92433 & $8.2\pm0.4$   & $5.4\pm0.3$   &  $6.9\pm0.1$  & $0.71\pm0.01$ & $10.84^{+0.19}_{-0.16}$ & $2.61^{+0.39}_{-0.35}$ \\
BR1202-0725N & SMG & 4.692 & 181.34575 &  -7.70825 & $19.4\pm0.5$  & $13.2\pm0.4$  &  $16.3\pm0.2$ & $1.76\pm0.02$ & $11.45^{+0.21}_{-0.26}$ & $2.87^{+0.43}_{-0.21}$ \\
BR1202-0725S & QSO & 4.695 & 181.34640 &  -7.70910 & $11.3\pm0.3$  & $7.7\pm0.2$   &  $16.1\pm0.2$ & $1.73\pm0.02$ & $11.46^{+0.45}_{-0.35}$ & $3.00^{+0.36}_{-0.61}$ \\
     BRI1335 & QSO & 4.407 & 204.51423 &  -4.54305 & $34.0\pm0.5$  & $21.0\pm0.3$  &  $20.2\pm0.2$ & $1.97\pm0.01$ & $12.05^{+0.30}_{-0.65}$ & $3.24^{+0.90}_{-0.90}$ \\
  J0331-0741 & QSO & 4.737 &  52.83193 &  -7.69533 & $9.4\pm0.5$   & $6.5\pm0.3$   &  $4.6\pm0.1$  & $0.51\pm0.02$ & $10.87^{+0.49}_{-0.46}$ & $1.88^{+0.29}_{-0.27}$ \\
  J0817+1351 & DLA & 4.260 & 124.42028 &  13.86062 & $6.9\pm0.4$   & $4.0\pm0.2$   &  $1.2\pm0.2$  & $0.11\pm0.01$ & $10.44^{+0.16}_{-0.14}$ & $1.95^{+0.48}_{-0.40}$ \\
 J0923+0247N & QSO & 4.655 & 140.76472 &  +2.79434 & $6.2\pm0.3$   & $4.2\pm0.2$   &  $3.1\pm0.1$  & $0.33\pm0.01$ & $11.09^{+0.58}_{-0.46}$ & $2.60^{+0.69}_{-0.74}$ \\
 J0923+0247S & SMG & 4.659 & 140.76369 &  +2.79321 & $4.1\pm0.3$   & $2.8\pm0.2$   &  $2.0\pm0.3$    & $0.22\pm0.03$ & $10.89^{+0.27}_{-0.29}$ & $1.53^{+0.77}_{-0.68}$ \\
  J1000+0234 & SMG & 4.538 & 150.22705 &  +2.57669 & $9.0\pm0.4$   & $5.8\pm0.3$   &  $8.1\pm0.1$  & $0.83\pm0.01$ & $10.81^{+0.17}_{-0.19}$ & $2.39^{+0.45}_{-0.29}$ \\
  J1341+0141 & QSO & 4.702 & 205.39249 &  +1.69936 & $12.8\pm0.6$  & $8.8\pm0.4$   &  $16.9\pm0.1$ & $1.83\pm0.01$ & $11.65^{+0.50}_{-0.66}$ & $2.95^{+0.60}_{-0.77}$ \\
J1511+0408N  & SMG & 4.685 & 227.98334 &  +4.13474 & $2.1\pm0.2$ & $1.5\pm0.1$ &  $1.9\pm0.1$  & $0.20\pm0.02$ & $11.04^{+0.22}_{-0.23}$ & $2.10^{+0.72}_{-0.44}$ \\
J1511+0408S  & QSO & 4.679 & 227.98323 &  +4.13418 & $8.3\pm0.3$   & $5.6\pm0.2$   &  $9.4\pm0.1$  & $1.00\pm0.01$ & $11.25^{+0.64}_{-0.36}$ & $2.95^{+0.13}_{-0.67}$ \\
SGP38326-1   & SMG & 4.425 &   0.78012 & -33.04738 & $16.6\pm0.3$  & $10.3\pm0.2$  &  $17.5\pm0.2$ & $1.72\pm0.02$ & $11.04^{+0.27}_{-0.36}$ & $2.52^{+0.28}_{-0.26}$ \\
SGP38326-2   & SMG & 4.430 &   0.77968 & -33.04755 & $4.3\pm0.2$   & $2.7\pm0.1$   &  $7.3\pm0.1$  & $0.72\pm0.01$ & $11.39^{+0.21}_{-0.22}$ & $2.44^{+0.56}_{-0.35}$ \\
\hline                                                                               
BR1202-0725-Ly$\alpha$ & - & 4.710 & 181.34602 & -7.70959 & $5.3\pm0.5$ & $3.6\pm0.3$ & $2.6\pm0.3$ & $0.28\pm0.03$ & $10.33^{+0.20}_{-0.22}$ & $1.98^{+0.47}_{-0.27}$ \\ 
SGP38326-3   & - & 4.438 & 0.78045   & -33.04793 & $1.2\pm0.2$ & $0.8\pm0.1$   & $1.2\pm0.1$  & $0.11\pm0.01$  & $10.63^{+0.25}_{-0.31}$ & $1.31^{+0.75}_{-0.66}$ \\
SGP38326-4 & - & 4.451 & 0.77987 & 33.04680 & $> 2.3$ & $> 1.4$ & $1.2\pm0.2$ & $0.12\pm0.02$ & $10.14^{+0.22}_{-0.20}$ & $1.59^{+0.55}_{-0.34}$ \\
\hline
\end{tabular}
}
\tablefoot{Column 1: adopted galaxy name. Column 2: original galaxy identification method. Column 3: galaxy redshift from the \cii\ line. Columns 4 and 5: galaxy kinematic center (from Paper\,III). Column 6 and 7: \cii\ flux and luminosity. Columns 8 and 9: 160-$\mu$m continuum flux and luminosity. Columns 9, 10, and 11: stellar mass, SFR averaged over the last 100 Myr from SED fitting (from Paper II).}
\end{table*}

Our selection gave us a sample of 14 sources with low-resolution \cii\ data: six galaxies (J0331-0741, J0923+0247N, J0923+0247S, J1341+0141, J1551+0408N, and J1551+0408S) from \citet{Trakhtenbrot2017}, three galaxies (AzTEC-159, J0817+1351, and J1000+0234) from \citet{Jones2017}, two galaxies (BR1202-0725N and BR1202-0725S) from \citet{Wagg2012}, and three more galaxies (ALESS-73.1, AzTEC-1, and BRI1335) from \citet{DeBreuck2014}, \citet{Tadaki2018}, and \citet{Wagg2010}, respectively. These objects were originally identified as submillimeter galaxies (SMGs), quasi-stellar objects (QSOs), and a metal-rich Damped Lyman-Alpha system (DLA). Four out of 14 galaxies had \cii\ data in the ALMA archive with angular resolutions of about $0.1''-0.2''$, so they were not reobserved. These objects are (i) our pilot galaxy ALESS-73.1 \citep{Lelli2021}, (ii) AzTEC-1 \citep{Tadaki2019}, (iii) BRI1335-0417 \citep{Tsuki2021}, and (iv) J0817+1351 \citep{Neeleman2020}. The last three galaxies have been reanalized by \citet{RomanOliveira2023} together with the galaxy group SGP38326 at $z\simeq4.4$. The two most luminous galaxies of this group (SGP38326-1 and SGP38326-2) satisfy the TRICEPS selection criteria, so they have been retroactively added to our sample, increasing it to 16 objects.

Half of the TRICEPS galaxies form a galaxy pair: BR1202-0725N \& BR1202-0725S; J0923+0247N \& J0923+0247S; J1511+0408N \& J1511+0408S; SGP38326-1 \& SGP38326-2. This allows comparing two equal-size sub-samples of galaxies in different ``environments'': objects with or without a nearby companion of similar mass. In addition, about half of the TRICEPS galaxies (7 out of 16) host an active galactic nucleus (AGN): the six QSOs and ALESS-73.1 \citep{Gilli2011}. For the other galaxies, there is no known evidence of the presence of an AGN.

\subsection{Sample properties}\label{sec:sampleprop}

The basic properties of the TRICEPS sample are summarized in Table\,\ref{tab:sample}. In addition to the 16 TRICEPS galaxies, Table\,\ref{tab:sample} provides the properties of three additional galaxies, which are well detected in \cii\ in the field of view of two different galaxy pairs: BR1202-0725 and SGP38326. The properties of these associations of galaxies are discussed in detail in Sect.\,\ref{sec:pairs}.

\begin{figure}[h]
\includegraphics[width=0.45\textwidth]{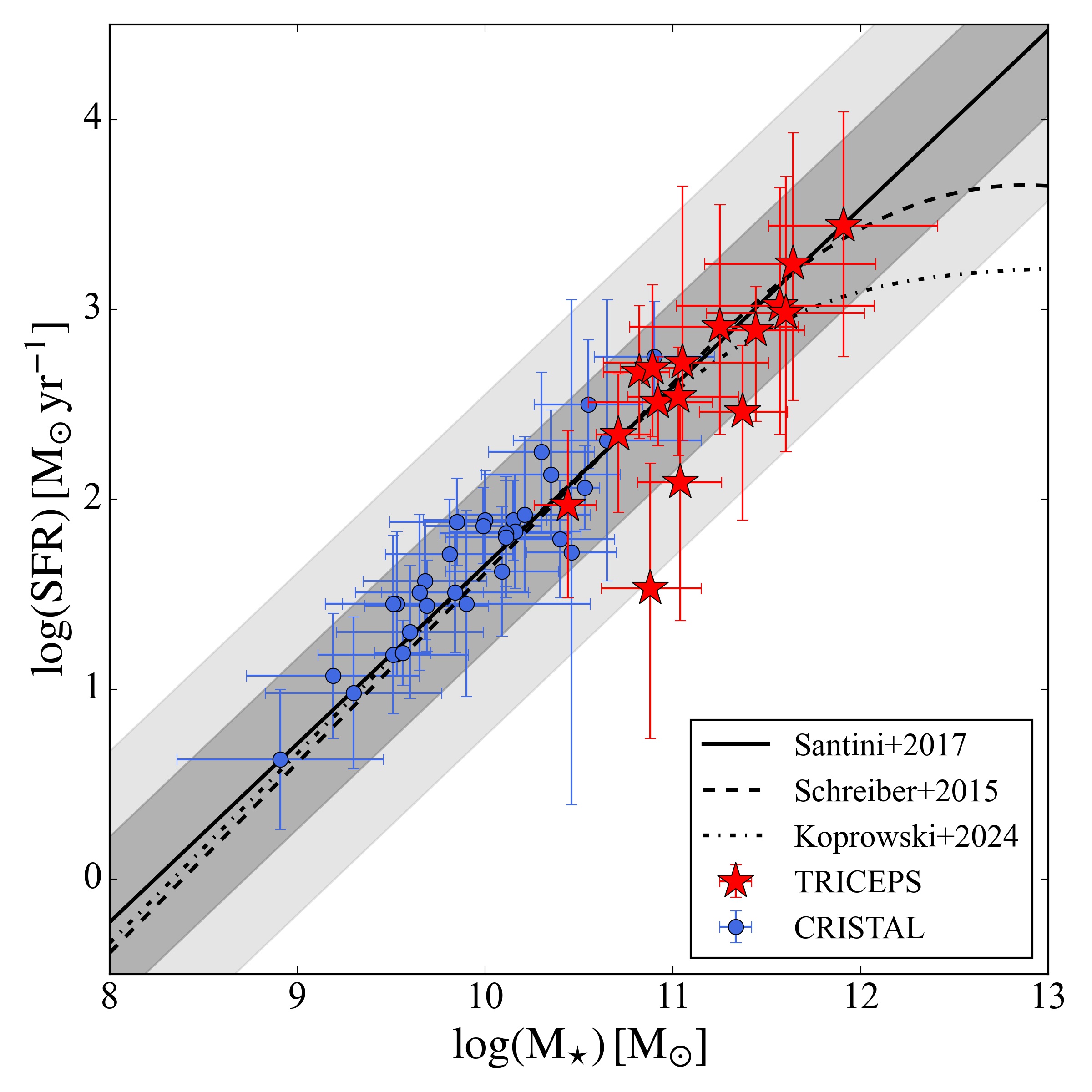}
\caption{The location of TRICEPS galaxies (red stars) on the $M_\star$-SFR plane. Blue circles show galaxies from the ALMA CRISTAL survey \citep{HerreraCamus2025A&A...699A..80H}. The solid black line shows the star-forming main sequence at $z\simeq4.5$ from \citet{Santini2017}; dark gray and light gray areas correspond to, respectively, 1$\sigma$ and 2$\sigma$ deviations from the mean, where $\sigma=0.45$ dex \citep{Santini2017}. The dashed and dotted lines show the non-linear relations from \citet{Schreiber2015} and \citet{Koprowski2024}, respectively. TRICEPS galaxies lie at the top-end of the main sequence, so they are representative of the most massive star-forming galaxies at this cosmic epoch.}
\label{fig:SFMS}
\end{figure}

Figure\,\ref{fig:SFMS} shows the location of TRICEPS galaxies on the $M_\star$-SFR plane. Measurements of $M_\star$ and SFRs come from SED fits using ALMA, JWST, and HST data, as described in Paper\,II. In particular, the AGN contribution is included in the SED fits, leading to relatively high uncertainties on $M_\star$ ($\sim$0.5 dex) and SFRs ($\sim$0.6 dex) for the QSOs (see Paper II). Nevertheless, TRICEPS galaxies clearly lie on the top end of the star-forming main sequence at $z\simeq4-5$, independently of whether we consider a linear relation \citep{Santini2017} or a non-linear one \citep{Schreiber2015, Koprowski2024}. TRICEPS galaxies have very high SFRs ($\sim$100$-$3000 M$_\odot$ yr$^{-1}$), so they are sometimes referred to as ``starbursts'' in the literature. However, contrary to the usual definition of ``starbursts'', they are not above the star-forming main sequence. Rather, TRICEPS galaxies are representative of the most massive star-forming objects at $4<z<5$, regardless of their original identification (SMG, QSO, DLA).

Figure\,\ref{fig:SFMS} includes galaxies from the ALMA-CRISTAL survey \citep{HerreraCamus2025A&A...699A..80H}, which targeted 19 galaxies at $4<z<6$. The CRISTAL sample subsequently increased to 39 galaxies because several targets turned out to be multiple systems, and nine galaxies with archival ALMA data were added. In total, 24 out of 39 galaxies ($\sim$60$\%$) in the CRISTAL sample belong to multiple systems. Figure\,\ref{fig:SFMS} shows that CRISTAL galaxies have systematically lower $M_\star$ and SFRs than TRICEPS galaxies, so the two surveys are highly complementary. TRICEPS has median $\langle M_\star \rangle \simeq 10^{11}$ M$_\odot$ and median $\langle \rm{SFR} \rangle \simeq277$ M$_\odot$ yr$^{-1}$, while CRISTAL has $\langle M_\star \rangle \simeq 10^{10}$ M$_\odot$ and $\langle\rm{SFR}\rangle\simeq58$ M$_\odot$ yr$^{-1}$. Given that (i) TRICEPS is probing more \cii\ luminous and more extended galaxies than CRISTAL, and (ii) TRICEPS has 2.5 times better angular resolution in \cii\ than CRISTAL (average beams of $0.12''$ and $0.29''$, respectively, see Sect.\,\ref{sec:imaging}), the TRICEPS galaxies are resolved with more resolution elements and higher signal-to-noise ratio (SNR). On average, the \cii\ emitting area of TRICEPS galaxies is resolved with 46 spatial resolution elements with SNR>3, making TRICEPS the largest sample of well-resolved galaxies at $z=4-5$ to date.

\begin{table}
\caption[]{Summary of ALMA observations. ToS is the time on source.}
\centering
\footnotesize
\label{tab:obs}
\begin{tabular}{lcc}
\hline
Target      & ToS [hrs] & ALMA ID \\
\hline
ALESS-73.1 & 1.5 & 2011.0.00124.S, 2017.1.01471.S \\ 
AzTEC-1     & 1.0 & 2015.1.01345.S, 2017.1.00127.S \\
AzTEC-159   & 2.1 & 2019.1.01587.S \\
BR1202-0725 & 1.3 & 2019.1.01587.S \\
BRI1335     & 3.8 & 2017.1.00394.S, 2018.1.01103.S \\
J0331-0741  & 1.5 & 2019.1.01587.S \\
J0817+1351  & 2.4 & 2015.1.01564.S, 2017.1.01052.S \\
J0923+0247  & 1.9 & 2019.1.01587.S \\  
J1000+0234  & 3.0 & 2019.1.01587.S \\ 
J1341+0141  & 1.3 & 2019.1.01587.S \\
J1511+0408  & 2.4 & 2019.1.01587.S, 2023.1.01724.S \\
SGP38326    & 4.5 & 2015.1.00330.S, 2018.1.00001.S \\
\hline
\end{tabular}
\end{table}

\begin{table*}
\caption[]{Properties of the \cii\ data and the continuum images.}
\centering
\footnotesize
\label{tab:cubes}
\renewcommand{\arraystretch}{1.3}
\resizebox{\textwidth}{!}{%
\begin{tabular}{lccccccccccc}
\hline
Target          & $\mathcal{R}$ & PSF$_{\cii}$           & PA$_{\cii}$ & $\sigma_{\rm cube}$ & $3\sigma_{\rm map}$ & $N_{\rm res, \cii}$ & PSF$_{\rm cont}$       & PA$_{\rm cont}$ & $\sigma_{\rm cont}$ & $N_{\rm res, cont}$\\
                &               & arcsec $\times$ arcsec & degree      & mJy beam$^{-1}$     & mJy beam$^{-1}$ \kms & & arcsec $\times$ arcsec & degree          & mJy beam$^{-1}$ & \\
\hline
ALESS-73.1   & 0.5 & $0.129 \times 0.105$ & $-63.6$ & 0.18 & 34 & 50  & $0.164\times0.128$ & $-56.9$ & 0.017 & 25\\
AzTEC-1      & 1.0 & $0.071 \times 0.063$ & $+85.5$ & 0.28 & 52 & 84  & $0.077\times0.068$ & $+87.1$ & 0.024 & 57\\
AzTEC-159    & 1.0 & $0.078 \times 0.075$ & $+9.6$  & 0.15 & 25 & 66  & $0.074\times0.072$ & $+23.2$ & 0.013 & 38\\
BR1202-0725N & 1.0 & $0.158 \times 0.110$ & $-67.9$ & 0.25 & 43 & 38  & $0.152\times0.106$ & $-67.3$ & 0.029 & 27\\
BR1202-0725S & 1.0 & $0.158 \times 0.110$ & $-67.9$ & 0.25 & 45 & 37  & $0.152\times0.106$ & $-67.3$ & 0.029 & 27\\
BRI1335      & 1.0 & $0.140 \times 0.101$ & $-67.7$ & 0.16 & 30 & 127 & $0.158\times0.128$ & $-78.7$ & 0.018 & 51\\
J0331-0741   & 1.0 & $0.065 \times 0.061$ & $+82.2$ & 0.28 & 50 & 28  & $0.062\times0.058$ & $+82.6$ & 0.019 & 17\\
J0817+1351   & 1.0 & $0.194 \times 0.142$ & $-54.7$ & 0.28 & 46 & 25  & $0.217\times0.145$ & $-50.6$ & 0.021 & 7\\
J0923+0247N  & 1.0 & $0.096 \times 0.083$ & $-68.4$ & 0.16 & 29 & 40  & $0.085\times0.077$ & $-80.6$ & 0.015 & 17\\
J0923+0247S  & 1.0 & $0.179 \times 0.162$ & $-58.1$ & 0.23 & 46 & 21  & $0.085\times0.077$ & $-80.6$ & 0.015 & 8\\
J1000+0234   & 1.0 & $0.138 \times 0.092$ & $-64.7$ & 0.16 & 32 & 38  & $0.134\times0.089$ & $-65.2$ & 0.015 & 28\\
J1341+0141   & 1.0 & $0.127 \times 0.088$ & $-69.8$ & 0.22 & 40 & 50  & $0.124\times0.085$ & $-69.3$ & 0.017 & 43\\
J1511+0408N  & 1.0 & $0.193 \times 0.158$ & $-73.6$ & 0.18 & 37 & 14  & $0.126\times0.085$ & $-66.9$ & 0.014 & 12\\
J1511+0408S  & 1.0 & $0.130 \times 0.090$ & $-67.4$ & 0.15 & 26 & 51  & $0.126\times0.085$ & $-66.9$ & 0.014 & 24\\
SGP38326-1   & 1.0 & $0.168 \times 0.151$ & $-82.8$ & 0.17 & 31 & 51  & $0.122\times0.112$ & $-77.8$ & 0.016 & 63\\
SGP38326-2   & 1.0 & $0.168 \times 0.151$ & $-82.8$ & 0.17 & 35 & 15  & $0.122\times0.112$ & $-77.8$ & 0.016 & 29\\
\hline
BR1202-0725-Ly$\alpha$ & 1.0 & $0.158 \times 0.110$ & $-67.9$ & 0.25 & 35 & 16  & $0.152\times0.106$ & $-67.3$ & 0.029 & 8 \\
SGP38326-3   & 1.0 & $0.168 \times 0.151$ & $-82.8$ & 0.17 & 26 & 8   & $0.122\times0.112$ & $-77.8$ & 0.016 & 4\\
SGP38326-4   & 1.0 & $0.168 \times 0.151$ & $-82.8$ & 0.17 & ... & ...   & $0.122\times0.112$ & $-77.8$ & 0.016 & $<5$ \\
\hline
\end{tabular}
}
\tablefoot{Column 1: galaxy name. Column 2: Briggs' robust parameter. Columns 3, 4, and 5: synthesized beam, position angle, and average rms noise of the \cii\ cube (for a velocity resolution of 30 \kms). Column 6: pseudo-3$\sigma$ value of the \cii\ intensity map. Column 7: number of resolution elements with SNR>3 across the \cii\ emitting area. Column 8, 9, and 10: synthesized beam, position angle, and rms of the continuum map. Column 11: number of resolution elements with SNR>3 across the continuum map.}
\end{table*}

\section{ALMA data}\label{sec:ALMA}

\subsection{New TRICEPS observations}\label{sec:newobs}

New ALMA band-7 observations were obtained for 10 galaxies using the 12\,m array in its configurations C43-6 or C43-7. Only a few observations were taken during ALMA Cycle 7 because of the telescope shutdown due to the Covid-19 pandemic. The majority of the data were taken during 2021$-$2022. For one target (J1511$+$0408), additional data were taken during 2024$-$2025 (project ID: 2023.1.01724.S, PI: F. Lelli).

Each galaxy was observed for $\sim$1$-$3 hours (see Table\,\ref{tab:obs}). Three galaxy pairs were observed with a single pointing, maximizing the survey efficiency. The pair members are typically separated by $2''-4''$, so they are well within the ALMA primary beam (with half-power beam-width of $\sim$17.5$''$); the pointing center was chosen in-between the two galaxies.

We used dual polarization and four spectral windows (SPWs) with a bandwidth of 1875 MHz each (about $1600-1700$ \kms) covered with 3840 channels. To reduce the data rate and have a line spread function (LSF) as close as possible to a Gaussian, we used spectral averaging over 8 channels with the default Hanning smoothing. This setup gives a LSF with a full-width half-maximum (FWHM) of 3.9 MHz (about $3.3-3.5$ \kms), which is effectively identical to the channel width (see Table 5.2 in the ALMA Technical Handbook). One SPW is centered on the redshifted \cii\ line, while the other three are used to trace the submm continuum and search for serendipitous line detections in the field of view. For galaxy pairs, we adjusted the central frequency to cover their \cii\ emission within a single SPW.

\subsection{Data reduction and imaging}\label{sec:imaging}

Data reduction was performed for all TRICEPS galaxies using the Common Astronomy Software Applications (\textsc{Casa}) package \citep{McMullin2007, CASA2022}. The Fourier-plane data was flagged and calibrated using the appropriate pipeline version provided by the ALMA observatory. The imaging was performed with the \texttt{tclean} task (version 6.5.4.9), using Briggs' weighting with robust parameter $\mathcal{R}=1.0$ and interactive cleaning. First, we created line$+$continuum cubes by imaging the SPW with the \cii\ line plus the adjacent continuum SPW at the native spectral resolution. Then, the continuum emission was subtracted in the image plane using the \texttt{imcontsub} task, fitting a first-order polynomial to the line-free channels. Given that our sources are spatially extended and not necessarily coincident with the ALMA pointing center (especially for the galaxy pairs), subtracting the continuum emission in the image plane is more robust than subtracting it in the $uv$ plane. In any case, subtracting the continuum emission in the $uv$ plane would not significantly change our results. We also created high-quality continuum images by combining all the line-free SPWs. In the rest of this paper, we use cubes and maps without primary-beam correction, so the root-mean-square (rms) noise is uniform and properly defined. The primary-beam maps are instead used to calculate correct total fluxes, but the flux correction is at most a few percent, smaller than the systematic uncertainty due to the overall flux calibration ($\sim$10$\%$).

To investigate the gas distribution and kinematics of galaxies, it is important to find the best possible compromise between spatial resolution, spectral resolution, and sensitivity. Such a combination can vary from galaxy to galaxy depending on observing conditions as well as galaxy properties (angular size, rotation velocity, inclination, and so on). For these reasons, our overall imaging strategy has a few exceptions: (i) for both the \cii\ cube and continuum map of ALESS-73.1, we use $\mathcal{R}=0.5$ to achieve adequate spatial resolution \citep{Lelli2021}; (ii) for the \cii\ cubes of J0923+0247S and J1511+0408N, we use a Gaussian taper of 0.1$''$ to increase the sensitivity at the expense of the spatial resolution; (iii) for the \cii\ cube and continuum map of the SGP38326, we use a Gaussian taper of 0.15$''$ and 0.10$''$, respectively. In future publications, we will consider different values of $\mathcal{R}$ and different $uv$ tapers to explore different angular scales. Here, we present ``reference'' cubes and maps that provide a good compromise between sensitivity, angular resolution, and shape of the point spread function (PSF) to study the gas kinematics.

To find the ideal compromise between spectral resolution and sensitivity, we spectrally smoothed the \cii\ cubes to lower resolutions, ranging from 2 to 12 times the native resolution ($3.3-3.5$ \kms). Given that the native LSF is well approximated by a Gaussian, we use Gaussian kernels so that the LSF of the smoothed cube is also a Gaussian. The smoothed cube is resampled to a channel width equal to the final FWHM of the LSF. After various trials, which are described in detail in Paper III, we find that a common velocity resolution of 30 \kms\ (FWHM) is a good compromise for each galaxy. Notably, if one bins the channels to 30 \kms\, before imaging in \textsc{Casa}, the resulting LSF will be the convolution of a Gaussian with a boxcar, which is less ideal to study the gas kinematics of galaxies, so we prefer to perform the spectral smoothing in the image plane.

Table\,\ref{tab:cubes} provides the properties of our reference continuum maps and \cii\ cubes. The continuum maps have a mean angular resolution of $0.11''$ with a standard deviation of $0.03''$. The \cii\ cubes have a mean angular resolution of $0.12''$ with a standard deviation of $0.03''$. For comparison, the Briggs-weighted \cii\ cubes of the CRISTAL survey \citep{HerreraCamus2025A&A...699A..80H} have a mean angular resolution of $0.29''$ with a standard deviation of $0.15''$. On average, therefore, the TRICEPS data have 2.5 times better angular resolution than the CRISTAL data, and are more homogeneous among themselves.

\subsection{$\cii$ moment maps, spectra, and fluxes}\label{sec:maps}

After imaging, the \cii\ cubes were analyzed using \bb\ \citep[][]{DiTeodoro2015}, in particular the Python-wrapper version 1.3.3.  We run the \bb\ source finder on all the \cii\ cubes to find all the detectable \cii\ emitters in the field of view and to create ``cubelets'' centered on each source, which are more manageable to analyze given their smaller size. Moment maps were created with the \texttt{Makemask} task, considering the signal inside a Boolean mask created with the \texttt{Smooth \& Search} option. When using a mask, the noise in the moment-zero map ($\sigma_{\rm map}$) varies from pixel to pixel. We built a SNR map considering channel dependencies \citep[Eq. A5 in][]{Lelli2014c}, then computed a pseudo 3$\sigma_{\rm map}$ value taking the median intensity of pixels with SNR between 2.75 and 3.25. Using the 3$\sigma_{\rm map}$ contour to define the area of the reliably detected \cii\ emission, the TRICEPS galaxies are resolved $-$ on average $-$ with 46 resolution elements (see Table\,\ref{tab:cubes}).

\begin{figure*}[h!]
    \centering
    \includegraphics[width=0.9\linewidth]{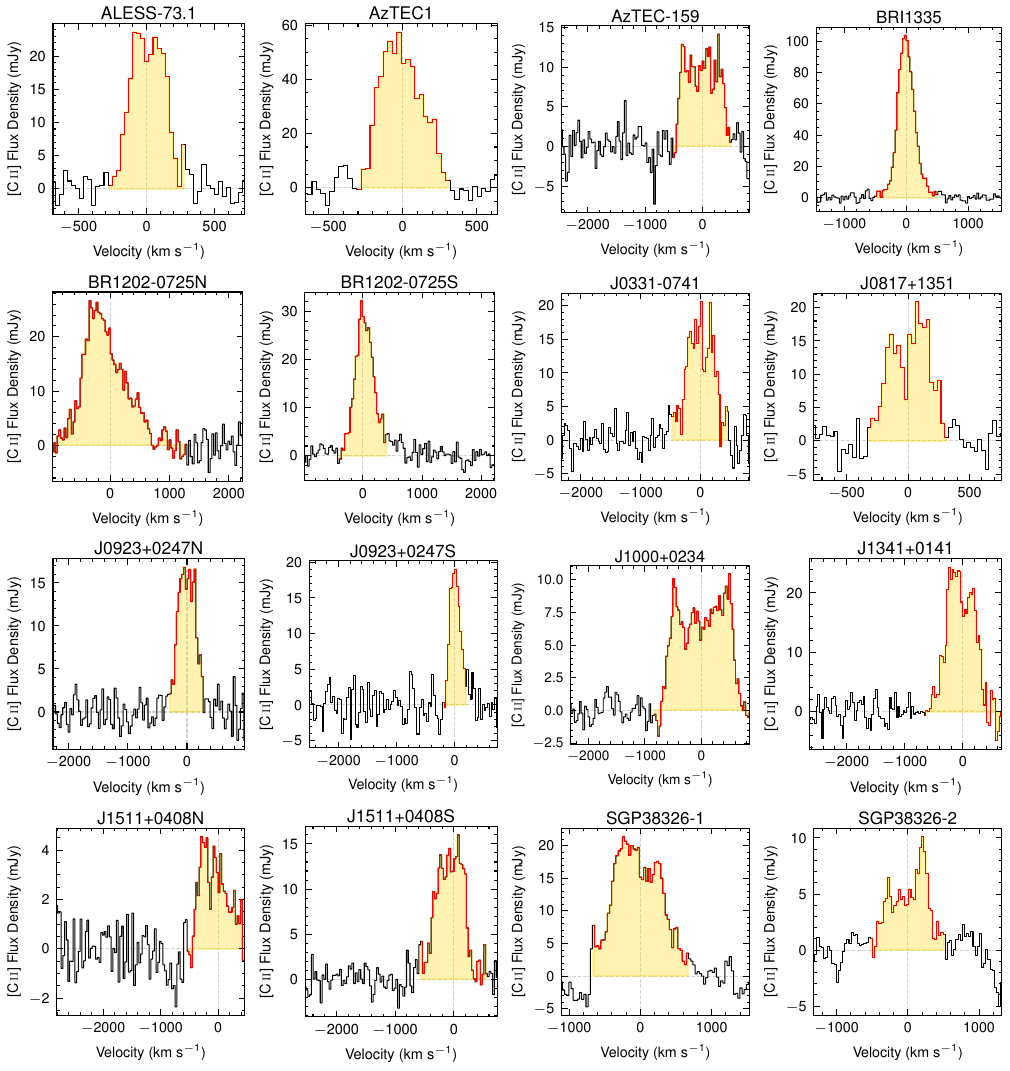}
    \caption{Spatially integrated \cii\ spectra. The red line and yellow area correspond to the velocity range used to calculate the \cii\ flux. The vertical dashed line corresponds to the galaxy redshift.}
    \label{fig:spectra}
\end{figure*}

We extracted global \cii\ spectra from the unmasked cubes using an aperture centered on the galaxy. Different apertures were tested. The optimal result was given by a radius of 4$R_{\rm e,\,\cii}$, where $R_{\rm e,\,\cii}$ is the effective radius from fitting a S\`ersic profile to the \cii\ moment-zero map. The S\'ersic fits will be presented in a future TRICEPS paper. For two galaxy pairs (J1511+0408N and J1511+0408S, SGP38326-1 and SGP38326-2), the aperture was visually defined to avoid contamination from the nearby companion. The \cii\ spectra were used to measure total \cii\ fluxes by summing the \cii\ emission within the appropriate spectral range. Fig.\,\ref{fig:spectra} shows the \cii\ spectra and the chosen integration ranges. Notably, the \cii\ spectra extracted with this procedure contain more flux than the moment-zero \cii\ maps because the masking procedure excludes pixels with $\rm{SNR}\lesssim2$, which however positively contribute to the total flux. For J1511+0408N, a small fraction ($\lesssim$10$\%$) of the \cii\ flux may be missing because the \cii\ emission is more spectrally extended than we expected, ending at the edge of the ALMA SPW.

We measured the rest-frame 160 $\mu$m flux summing the pixels within a circular aperture of 4$R_{\rm e, cont}$, where $R_{\rm e, cont}$ is the effective radius from fitting a S\`ersic profile to the continuum map. Paper II provides alternative measurements of the continuum flux for the purpose of SED fitting, after smoothing the images to the same angular resolution of our JWST/MIRI images. The two flux measurements are compatible.

\begin{figure}[h]
\includegraphics[width=0.45\textwidth]{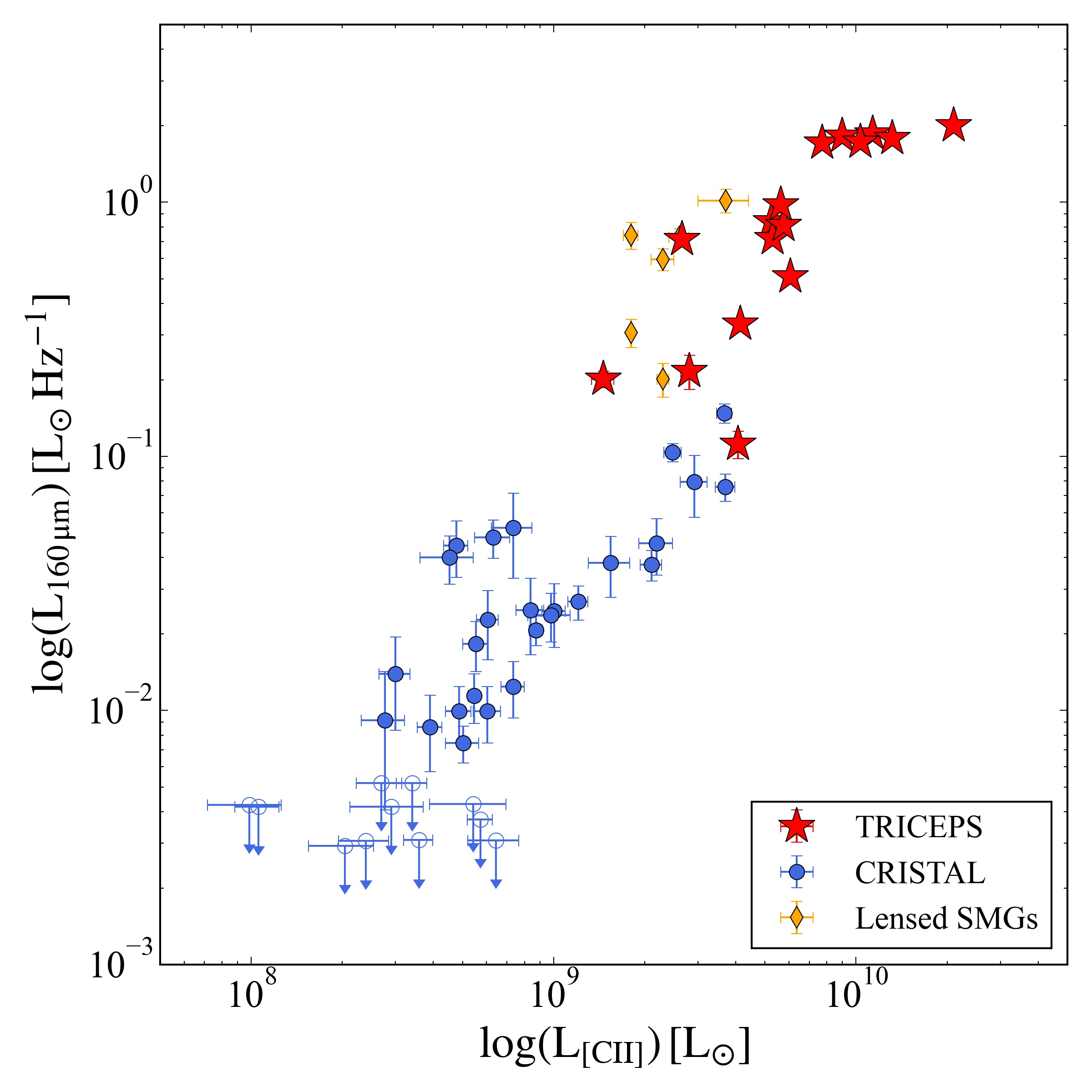}
\caption{The location of TRICEPS galaxies (red stars) on the $L_{\rm{[CII]}}-L_{160\mu m}$ plane. Orange diamonds show gravitationally lensed SMGs \citep{Reuter2020ApJ...902...78R, Rizzo2021}, while blue circles show galaxies from the CRISTAL survey \citep{HerreraCamus2025A&A...699A..80H}. Empty circles represent CRISTAL galaxies for which the 160 $\mu$m continuum is undetected.}
\label{fig:CIIvsDust}
\end{figure}

Figure \ref{fig:CIIvsDust} shows $L_{\rm [CII]}$ versus $L_{160\mu m}$ for TRICEPS galaxies in comparison to CRISTAL galaxies \citep{HerreraCamus2025A&A...699A..80H} and a small sample of gravitationally lensed SMGs at $z\simeq4-5$ with high-resolution \cii\ data \citep{Reuter2020ApJ...902...78R, Rizzo2021}. As expected, there is a broad trend between $L_{\rm [CII]}$ and $L_{160\mu m}$. TRICEPS and CRISTAL galaxies lie at the top and bottom ends of this trend, respectively, indicating again the large complementarity between the two samples. The lensed SMGs cover a similar location as the TRICEPS galaxies. If we interpret $L_{\rm [CII]}$ as a tracer of the cold gas mass and $L_{160\mu m}$ as a tracer of the SFR, it is evident that TRICEPS galaxies are among the most gas-rich and most star-forming objects at $z\simeq4-5$.

\begin{figure*}
\centering
\includegraphics[width=0.9\textwidth]{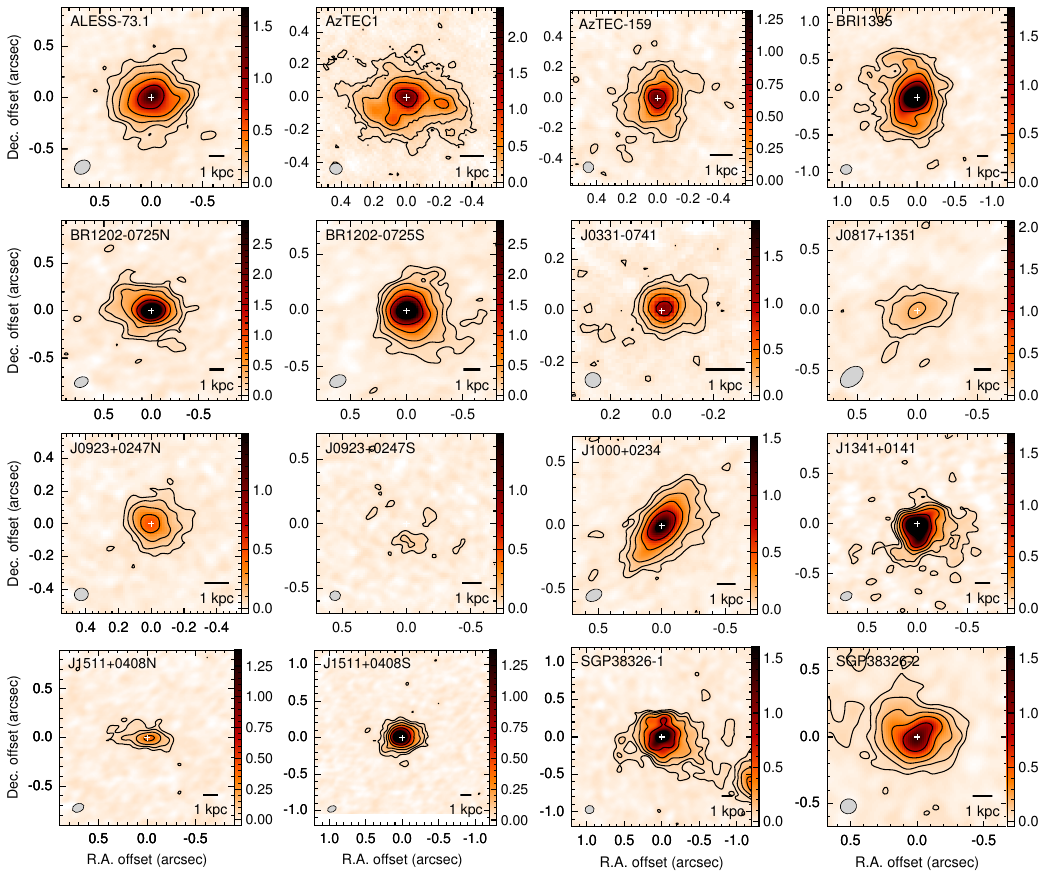}
\caption{Continuum maps at restframe 160 $\mu$m of TRICEPS galaxies. The colorbar is in units of mJy/beam. Isodensity contours are at 3, 6, 12, 24 $\times$ $\sigma_{\rm cont}$ (see Table\,\ref{tab:cubes}). The star shows the galaxy center. The grey ellipse shows the ALMA synthesized beam. The bar correponds to 1 kpc.}
\label{fig:cont}
\end{figure*}

\begin{figure*}
\centering
\includegraphics[width=0.9\textwidth]{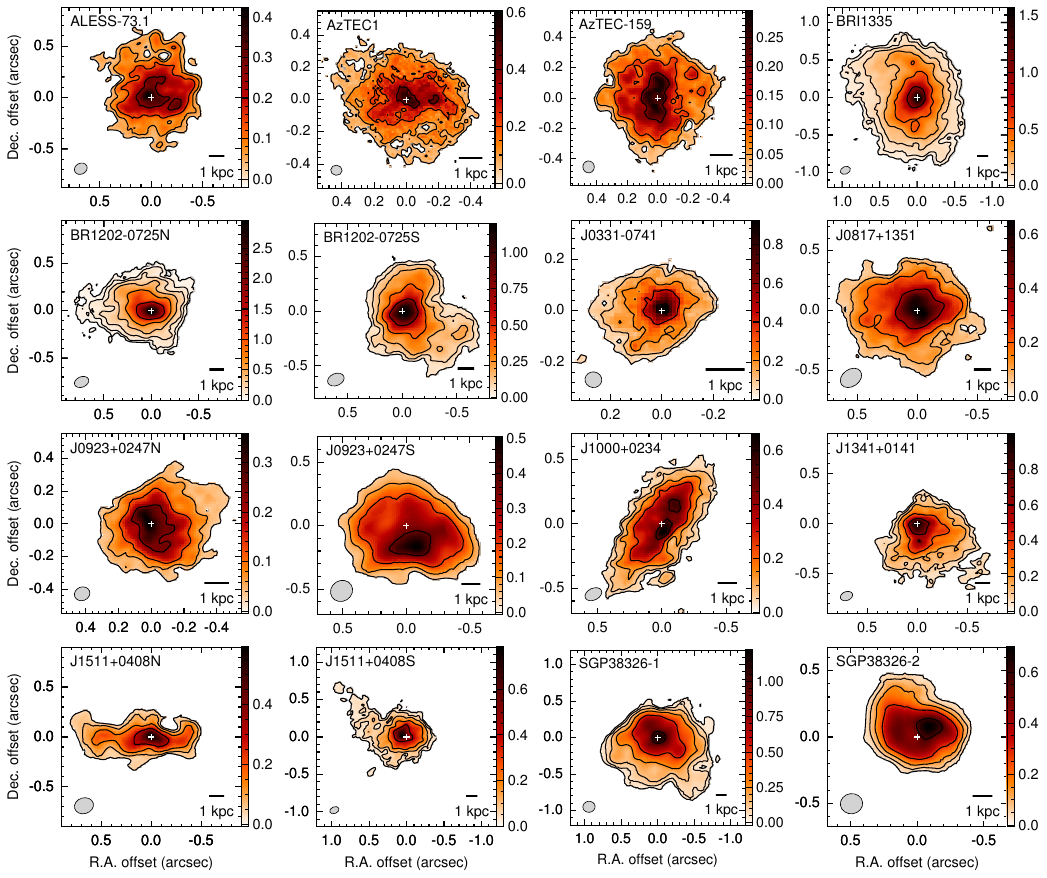}
\caption{\cii\ intensity maps (moment zero) of TRICEPS galaxies. The colorbar is in units of Jy beam$^{-1}$ \kms. Contours are at $1,2,4,8...\times3\sigma_{\rm cont}$ (see Table\,\ref{tab:cubes}). The other symbols have the same meaning as in Fig.\,\ref{fig:cont}.}
\label{fig:mom0}
\end{figure*}

\section{Results}\label{sec:results}

\subsection{Dust and gas morphology}\label{sec:gasdist}

Figure\,\ref{fig:cont} shows the 160-$\mu$m rest-frame continuum maps of TRICEPS galaxies tracing the dust distribution, while Figure\,\ref{fig:mom0} shows the \cii\ intensity maps (moment zero) tracing the cold gas distribution. In all galaxies, the dust and \cii\ emissions are well detected and resolved with tens of resolution elements, up to 125 in the \cii\ map of BRI1335. The only exception is the dust continuum emission of J0923+0247S, which is only marginally detected at 3$\sigma$. The gas and dust distributions are relatively smooth at our spatial resolutions of $\sim0.4-1.3$ kpc.

Importantly, the gas distribution does not show massive kpc-size clumps, which have been studied in rest-frame UV and optical images of star-forming galaxies at $z\simeq2-6$ \citep[e.g.,][]{ForsterSchreiber2011, Guo2012, Ribeiro2017}. Clearly, the \cii\ maps are not perfectly symmetric, nor entirely free of substructures, but appear quite similar to the \hi\ and CO maps of local star-forming galaxies observed at similar physical resolutions (see the \hi\ atlases in \citealt{Verheijen2001, Noordermeer2005} and the CO atlases in \citealt{Bolatto2017, Lin2020}). In other words, the gas morphology on sub-kpc scales in massive galaxies $z\simeq4-5$ does not appear peculiar with respect to that in massive galaxies at $z\simeq0$. In a future TRICEPS paper, we will study possible sub-structures by subtracting a smooth, axisymmetric, beam-convolved model to the continuum map, as we did in \citet{Lelli2021} for our pilot galaxy ALESS-73.1.

The gas emission is more extended than the dust emission, as can be assessed by eye from the galaxy atlas in the Appendix. This phenomenology is a common characteristic of galaxies at $z>4$ \citep[e.g.,][]{Ginolfi2020, Fujimoto2020}, as well as galaxies at $z\simeq0$ when considering the \hi\ distribution \citep[e.g.,][]{Kennicutt2012}. The relative extension of gas and dust distributions will be quantified in a future TRICEPS paper. We anticipate that the average ratio between the \cii\ and the dust effective radii is around 1.5.
\begin{figure*}
\centering
\includegraphics[width=0.9\textwidth]{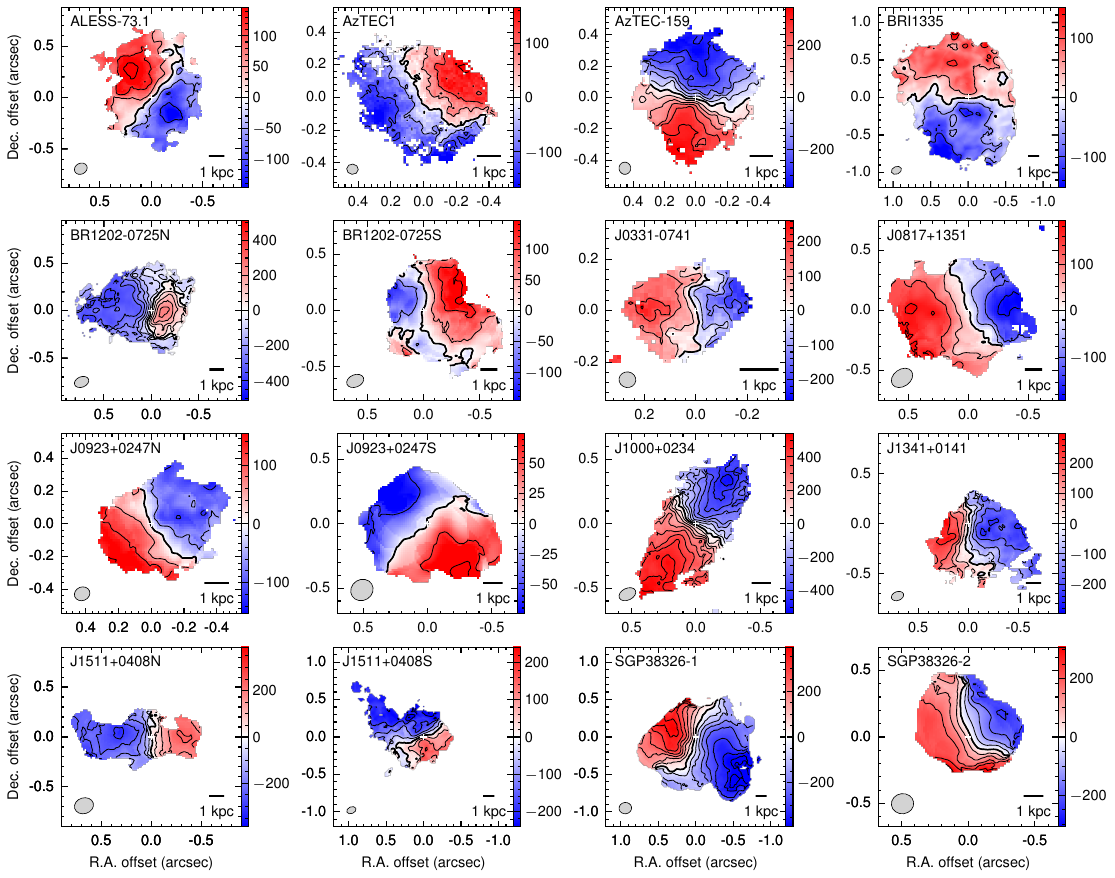}
\caption{\cii\ velocity maps (moment one) of TRICEPS galaxies. The colorbar is in units of \kms. The thick contour correspond to the mean redshift; the other contours are at $\pm60,\pm120,\pm180...\rm km\,s^{-1}$. The other symbols have the same meaning as in Fig.\,\ref{fig:cont}.}
\label{fig:mom1}
\end{figure*}

\begin{figure*}
\centering
\includegraphics[width=0.9\textwidth]{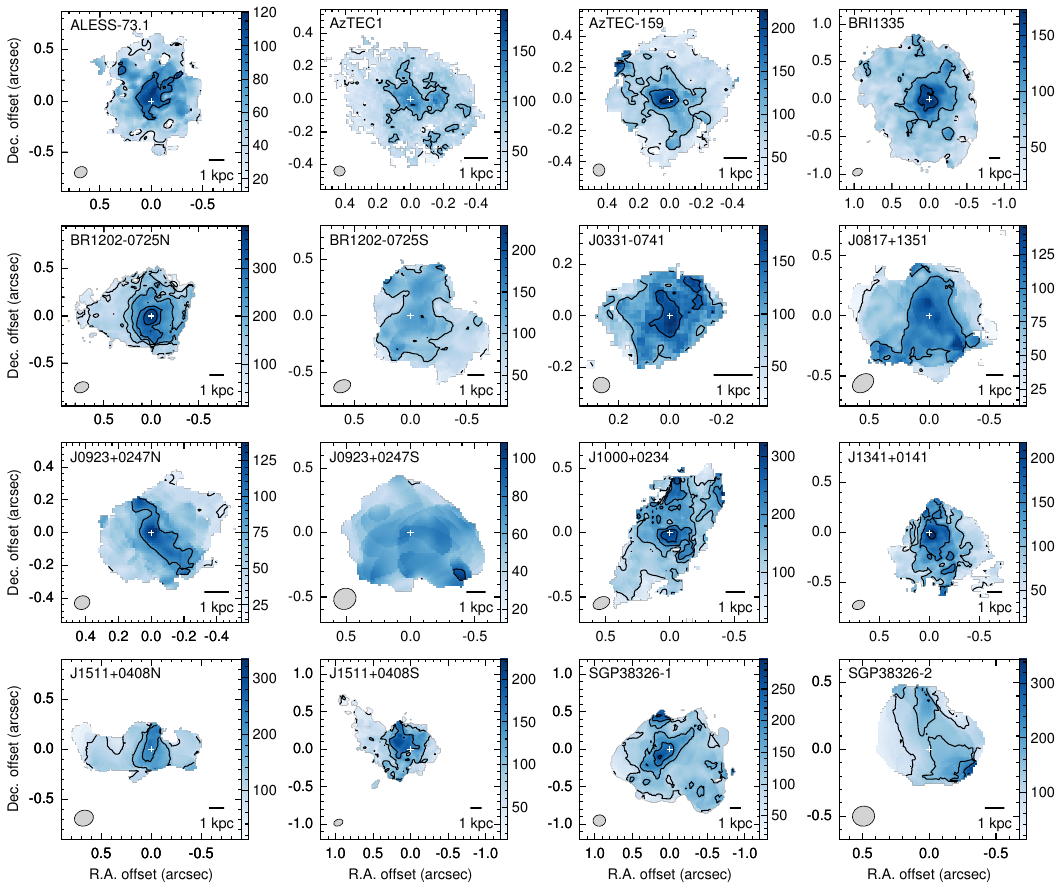}
\caption{\cii\ line-broadening maps (moment two) of TRICEPS galaxies. The colorbar is in units of \kms. Contours are at $+30, +90, +150...\rm km\,s^{-1}$ (in steps of of 60 \kms). The other symbols have the same meaning as in Fig.\,\ref{fig:cont}.}
\label{fig:mom2}
\end{figure*}

\subsection{Gas kinematics}\label{sec:gaskin}

Figure\,\ref{fig:mom1} shows the \cii\ velocity maps (moment one) of TRICEPS galaxies. In all galaxies, the \cii\ velocity maps display the typical characteristics of a rotating disk, suggesting that gas rotation is ubiquitous in massive star-forming galaxies at $z\simeq4-5$. For low-resolution emission-line data, velocity maps can sometimes be difficult to interpret because rotating disks, gas outflows, and galaxy mergers may produce similar beam-smeared velocity gradients \citep[e.g.,][]{Rizzo2022}. This is not the case for TRICEPS galaxies because the \cii\ emission is resolved with plenty of resolution elements (see Table \ref{tab:cubes}) and most velocity maps display the typical ``spider diagram'' of a rotating disk. Indeed, when a thin (approximately 2D), co-planar, rotating structure is projected on the sky, the resulting velocity map has several unique characteristics: (i) the velocity gradient is aligned with the geometric major axis; (ii) the systemic velocity (the mean redshift) forms a straight line aligned with the geometric minor axis (unless there are strong radial motions); and (iii) positive and negative isovelocity contours are symmetric with respect to the center, and V-shaped if the rotation curve flattens out. In general, these characteristics are not expected for gas outflows and galaxy mergers because these are intrinsically 3D structures with complex kinematics: at each position on the sky, multiple velocity vectors are projected along the line of sight, leading to broad and multi-peaked line profiles, so we expect much more complex velocity maps. The latter ones are not expected to resemble a spider diagram when observed with adequate spatial resolution. In other words, for a nearly 2D structure it is sensible to assign a single, physically meaningful, line-of-sight velocity at each position on the sky, whereas this is not true for a 3D structure in which multiple velocity components are projected at each sky position. 

For 10 out of 16 galaxies (ALESS-73.1, AzTEC-1, AzTEC-159, BRI1335, J0331-0741, J0817+1351, J1341+0141, J0923+0247N, J0923+0247S, J1000+0234), the velocity maps are quite regular and symmetric, indicating that non-circular motions are subdominant with respect to circular ones. This applies to both SMGs and QSOs with no clear distinction between them, suggesting that AGN feedback is not strongly affecting the overall \cii\ kinematics (see also Paper III). For the remaining six cases, the velocity maps show some distortions and/or irregularities. These galaxies form three pairs (BR1202-0725N \& BR1202-0725S, J1511+0408N \& J1511+0408S, SGP38326-1 \& SGP38326-2), which show clear signs of tidal interactions, as we discuss in Sect.\,\ref{sec:pairs}. Thus, it is likely that the kinematic irregularities are due to non-circular motions driven by the mutual interaction. In Paper III, we present full 3D modeling of the \cii\ cubes \citep[similarly to][for ALESS\,73.1]{Lelli2021}, showing that a rotating disk model provides a good representation of the gas kinematics in all galaxies, even the less regular ones.

Figure\,\ref{fig:mom2} shows the \cii\ line broadening maps (moment two) of TRICEPS galaxies. In most cases, the \cii\ line is maximally broad near the galaxy center, as expected for a rotating disk due to beam-smearing effects. We stress that the moment-two maps do not represent the intrinsic gas velocity dispersion (due to thermal and turbulent motions) because of the effects of spatial and spectral resolution. Full 3D modeling of the \cii\ cubes is needed to measure the intrinsic gas velocity dispersion \citep[e.g.,][]{DiTeodoro2015} and presented in Paper III.

\subsection{Galaxy pairs and tidal interactions}\label{sec:pairs}

\begin{figure*}
\centering
\includegraphics[width=0.9\textwidth]{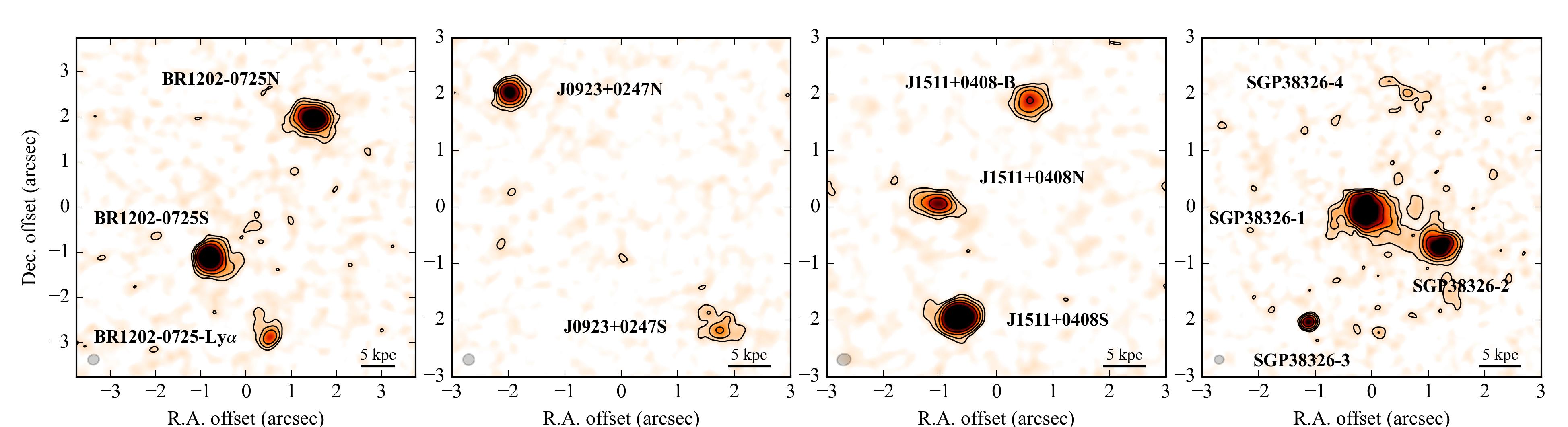}
\includegraphics[width=0.9\textwidth]{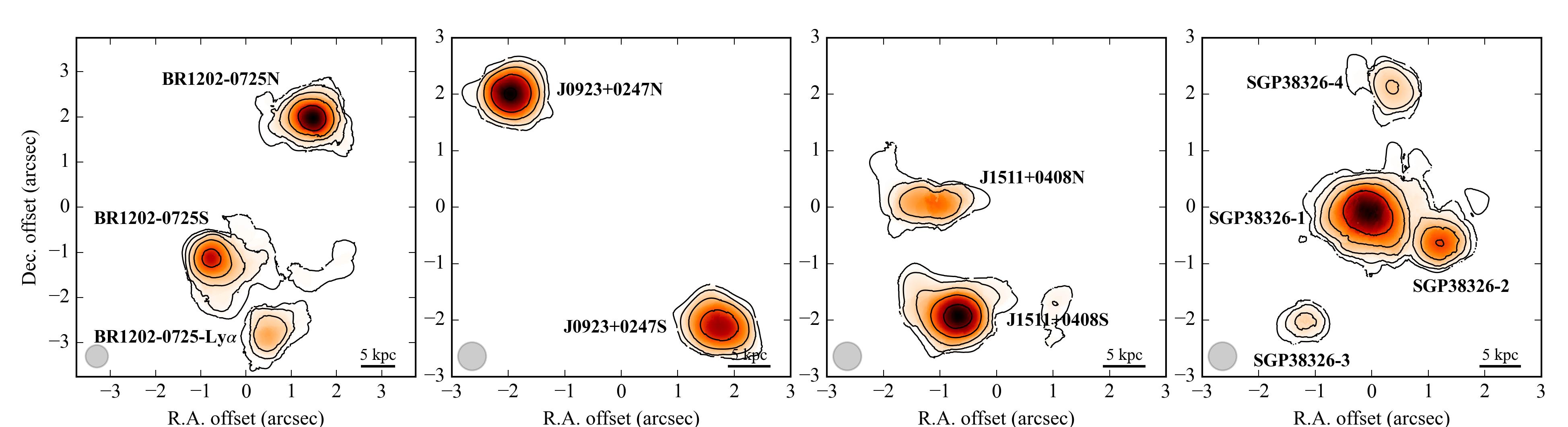}
\caption{High-sensitivity continuum maps (top panels) and \cii\ moment-zero maps (bottom panels) for galaxies in pairs. The labels indicate the galaxy name. The other symbols have the same meaning as in Fig.\,\ref{fig:cont}. See Sect.\,\ref{sec:pairs} for details.}
\label{fig:groups}
\end{figure*}

The TRICEPS sample contains four galaxy pairs, which are likely residing in the central regions of massive protoclusters, as we show in Paper IV. These overdensities are characterized by either a dominant QSO-SMG pair (BR1202-0725, J0923+0247, J1511+0408) or a dominant SMG-SMG pair (SGP38326). Figure\,\ref{fig:groups} shows deep continuum maps (top panels) and \cii\ moment-zero maps (bottom panels) of these four galaxy pairs. To maximize sensitivity to low surface brightness emission, the continuum maps were obtained using a Gaussian $uv$ taper of 0.15$''$, while the \cii\ moment maps using $\mathcal{R}=2$ and a restoring beam of $0.5"$. A detailed study of the diffuse \cii\ emission of all TRICEPS galaxies will be presented in a future paper.

The galaxy pair BR1202-0725 has been previously studied by \citet{Carilli2013} and \citet{Carniani2013} using ALMA Cycle-0 data at low resolution. The QSO to the South (BR1202-0725S) and the SMG to the North (BR1202-0725N) are clearly interacting, being connected by a \cii\ bridge \citep[see Fig. 1 in][]{Carilli2013}. This bridge is only partially visible in our data due to their higher resolution and lower sensitivity. The continuum and \cii\ maps of these two galaxies are significantly lopsided (Fig.\,\ref{fig:cont} and Fig.\,\ref{fig:mom0}), which is probably the result of a tidal interaction. In addition, the system contains two Ly$\alpha$ emitters, which have been associated with \cii\ emission \citep{Carilli2013}. Our new observations do not clearly detect Ly$\alpha$1, but we do detect Ly$\alpha$2 in both \cii\ and continuum emission, so we rename it as BR1202-0725-Ly$\alpha$ for simplicity. BR1202-0725-Ly$\alpha$ shows a lopsided dust distribution with a small tail in the direction of the QSO, probably due to tidal interactions. BR1202-0725-Ly$\alpha$ has $I_{\cii} \simeq 5.3$ Jy \kms, above the TRICEPS flux-based selection (see Sect.\,\ref{sec:sample}), but we do not consider it to be part of the proper TRICEPS sample because the \cii\ emission is resolved only with a few resolution elements with low SNR. We consider this galaxy part of an extended TRICEPS sample.

The galaxy pairs J0923+0247 and J1511+0408 have been previously studied by \citet{Trakhtenbrot2017} using ALMA Cycle-2 data at low resolution. In the former group, the QSO to the North (J0923+0247N) and the SMG to the South (J0923+0247S) are at a projected separation of $\sim$30 kpc; both the continuum and \cii\ maps show no sign of interaction. In the latter group, the SMG to the North (J1511+0408N) and the QSO to the South (J1511+0408S) are at a projected separation of $\sim$15 kpc; the \cii\ maps of both galaxies show tail-like extensions, which are likely due to a tidal interaction. The J1511+0408 system displays a third continuum source to the North, named J1511+0408-B following \citet{Trakhtenbrot2017}. This third galaxy is not detected in \cii\ emission, so it is unclear whether it is associated to the pair or is in the background/foreground.

The galaxy pair SGP38326 has previously been studied by \citet{RomanOliveira2023} using the same ALMA data. Our deep continuum and \cii\ maps (Fig.\,\ref{fig:groups}) reveal that the two SMGs (SGP38326-1 and SGP38326-2) are clearly interacting: they are connected by a dust bridge and embedded in a common \cii\ envelope. In addition to the SMG-SMG pair, there are two additional continuum sources that are also detected in \cii\ emission. The one to the North (SGP38326-4) is probably interacting with the main pair, but unfortunately its \cii\ emission lies at the edge of the observed bandwidth, so we cannot derive reliable \cii\ maps and measure precise physical quantities, such as \cii\ flux, redshift, and so on. Similarly to BR1202-0725-Ly$\alpha$, both SGP38326-3 and SGP38326-4 are considered part of the extended TRICEPS sample.

An additional proto-group or proto-cluster is represented by the system J1000+0234. Recent JWST/NIRSpec observations reveal several companion galaxies around the central SMG \citep{Solimano2024a}. In addition, low-resolution, high-sensitivity ALMA data \citep{Solimano2024b} reveal a 15-kpc long tail that extends beyond the rotating \cii\ disk. This gaseous tail is not seen in our higher-resolution, lower-sensitivity data. The tail morphology is similar to those seen in \hi\ emission in galaxies at $z\simeq0$ \citep[e.g.,][]{Lelli2012a, Lelli2014c, Lelli2015} and may be due to tidal interactions or cold gas accretion.

\section{Discussion}

\subsection{Ubiquitous rotation in massive galaxies at $z=4-5$}\label{sec:rotation}

In the previous sections, we presented dust continuum maps and \cii\ moment maps at sub-kpc spatial resolutions ($\sim$0.8 kpc on average) for 16 massive ($M_\star\simeq10^{10.5}-10^{12}$ M$_\odot$) galaxies at $z=4-5$, when the Universe was less than 1.5 Gyr old. The gas and dust distributions are relatively smooth, without clear massive clumps on kpc scales (Sect.\,\ref{sec:gasdist}). Remarkably, rotating gas disks are ubiquitous in massive star-forming galaxies at this cosmic epoch (Sect.\,\ref{sec:gaskin}). This holds independently of whether the galaxy hosts an AGN or not, implying that AGN feedback does not disrupt the overall gas rotation and must have (at most) a localized effect on the cold gas kinematics. Rotating gas disks are present even in galaxy pairs that are undergoing tidal interactions (Sect.\,\ref{sec:pairs}), albeit they are somewhat more irregular and disturbed than those of ``isolated'' galaxies.

Galaxies of comparable stellar mass at $z=0$ display the same phenomenology. Naturally, rotating gas disks are ubiquitous in massive star-forming spirals \citep[e.g.,][]{Noordermeer2005, DiTeodoro2021, DiTeodoro2023}, but they are also common in passive galaxies (ellipticals and lenticulars) that retained a cold gas component \citep[e.g.,][]{Serra2012, Davis2013, Davis2016, Shelest2020}. In addition, rotating gas disks are preserved during galaxy interactions at $z\simeq0$ \citep{Hibbard1996, Perna2022}, although they may be perturbed by tidal forces \citep[see, e.g., the iconic Arp\,220 major merger,][]{Scoville2017}. The rotating gas disk is eventually destroyed only if a full coalescence takes place \citep[see, e.g., the late-state merger prototype NGC\,7252,][]{Lelli2015}, but it could also reform after the merger is fully completed \citep{Ueda2014}. From this kinematic perspective, massive galaxies at $z=4-5$ do not appear particularly different from galaxies at $z=0$.

At lower stellar masses ($M_\star\lesssim10^{10.5}$ M$_\odot$), the observational situation is less clear and has been steadily changing over time with improvements in data quality and modeling techniques. The ALPINE team performed a visual classification of 75 star-forming galaxies at $4<z<6$ with \cii\ data at a mean resolution of $0.1''$ ($\sim$6 kpc), concluding that they show a wide variety of kinematic behaviors \citep{LeFevre2020}. Considering only galaxies with SNR$>$5, they found that 51$\%$ are mergers, 32$\%$ are extended dispersion-dominated objects, and only 17$\%$ are rotating disks. Subsequently, the ALPINE team focused on a subsample of 29 galaxies with their best \cii\ data (corresponding to $M_\star \simeq 10^{9.5}-10^{11}$ M$_\odot$) and performed a more quantitative analysis \citep{Jones2021}, concluding that about half of the galaxies cannot be kinematically classified, while the other half consists of 6/14 (43$\%$) rotating disks, 5/14 (36$\%$ mergers), and 3/14 (21$\%$) dispersion-dominated objects. The CRISTAL survey performed \cii\ follow-up observations of the 32 ALPINE galaxies in the same mass range ($M_\star \simeq 10^{9.5}-10^{11}$ M$_\odot$) with a mean resolution of $0.29''$ ($\sim$2 kpc), concluding that 50$\%$ of them are disks, while the remaining 50$\%$ are labeled as ``non disks'' \citep{Lee2025A&A...701A.260L}. The general trend is that better kinematic data and better modeling techniques led to a steady increase in the fraction of rotating disks, and a corresponding decrease in that of mergers and pressure-supported objects.

The previous trend should not come as a surprise because there is a long history of observational biases in studying galaxy kinematics at different $z$. In the nearby Universe ($z\simeq0$), the first studies of low-mass galaxies (baryonic masses $M_{\rm bar} \lesssim 10^8$ M$_\odot$) using resolved \hi\ observations concluded that their gas content is dominated by chaotic motions rather than by rotation \citep{Sargent1986, Lo1993}. Subsequent studies with improved observations lead to the opposite picture: low-mass star-forming galaxies can possess rotating gas disks down to $M_{\rm bar} \simeq 10^6$ M$_\odot$ and rotation velocities of $\sim$10-15 \kms \citep{McNichols2016, Iorio2017, Lelli2022}. Moving to cosmic noon ($z\simeq1-3$), studies using the first H$\alpha$ and \oiii\ integral-field spectroscopic surveys of massive galaxies suggested that 1/3 of them were rotation-dominated disks, 1/3 dispersion-dominated objects, and another 1/3 merging systems \citep{Forster2009, Gnerucci2011}. Subsequent studies with improved observations lead to a drastically different picture: at least 80$\%$ of star-forming galaxies at $z\simeq1-3$ have rotating gas disks, whereas dispersion-dominated and merging systems constitute a minority of the star-forming population \citep{Wisnioski2015, Wisnioski2019, Stott2016}. It is therefore possible that the diversity of kinematic behaviors found at $z=4-6$ by ALPINE and CRISTAL is not a true property of primeval galaxies, but it is driven by data quality. 

Broadly speaking, limited angular resolutions, limited spectral resolutions, and limited sensitivities have the general effect of ``hiding'' rotating disks and making them look like more complex kinematic structures. The limited angular resolution can make a rotating disk look like a pressure-supported object when resolved with a small number of resolution elements due to beam-smearing effects \citep[see, e.g., Figure 6 in][]{DiTeodoro2016}. The limited spectral resolution has a similar effect in low-inclination and/or low-mass disks because they display small line-of-sight velocity gradients, which require adequate velocity resolution to be identified. The limited sensitivity leads to a variety of effects, such as creating complex moment maps due to noise peaks and/or recovering only a fraction of a rotating disk due to intrinsic asymmetries in the gas distribution. The opposite effect is also possible: a galaxy merger may look like a disk when observed at low spatial and spectral resolution \citep{Rizzo2022}. However, this occurs only under very specific circumstances: the two galaxies must have a comparable gas mass (or comparable flux of the specific line under investigation) and be close to each other on the sky as well as in line-of-sight velocity, but without their disks overlapping too much both in space and velocity, otherwise they cannot mimic a relatively symmetric velocity field and/or major-axis position-velocity diagram. Clearly, this is a rarer effect than the general effects on ``isolated'' rotating gas disks. In any case, to clarify the observational situation in low-mass galaxies at $4<z<6$, the only solution is to obtain observations with higher resolution and sensitivity, possibly by upgrading ALMA with additional antennas on both short and long baselines (the proposed ALMA-2040 project).

\subsection{Future TRICEPS science}\label{sec:prospects}

The TRICEPS project has several scientific goals. In addition to the companion papers introduced in Sect.\,\ref{sec:intro}, we plan to perform the following studies in the near future:
\begin{itemize}
    \item Characterization of the gas, dust, and stellar distribution at sub-kpc scales, including measurements of surface brightness profiles, sizes, and sub-structures;
    \item Spatially resolved SED fitting at sub-kpc scales to derive stellar mass profiles and SFR profiles;
    \item Construction of mass models to constrain the amount and distribution of baryons and dark matter;
    \item Detailed investigation of possible non-circular motions in the gaseous disk, sub-dominant with respect to the regular rotation, and their potential connection to stellar bars, spiral arms, or galaxy interactions;
    \item Measurements of the angular momentum content of gas disks and its relation with cosmological gas accretion;
    \item Characterization of the extended \cii\ emission at large radii by smoothing and/or stacking the high-resolution data;
    \item Investigation of various SF laws beyond the ``classical'' Kennicutt law, such as relations considering orbital times \citep{Kennicutt1998}, volume densities \citep{Bacchini2019}, and gas pressure \citep{Ostriker2022};
    \item Investigation of disk stability using the classic $Q$ parameter \citep{Toomre1964} and its 3D version \citep{Bacchini2024};
    \item Comparisons of the observations with hydrodynamic simulations of galaxy formation by using mock-observed simulated galaxies \citep[e.g.,][]{Marasco2025}.
\end{itemize}
The current paper, therefore, provides only a glimpse of the exciting science that the TRICEPS dataset will make possible.

\section{Conclusions}\label{sec:summary}

We introduced the TRICEPS survey: a sample of 16 massive galaxies at $z\simeq4-5$ with (i) high-resolution ALMA data of the \cii\ line and rest-frame 160 $\mu$m continuum emission, (ii) multi-band JWST images with NIRCam and MIRI, and (iii) multi-line CO observations from ALMA and the VLA. In this first paper of the TRICEPS series, we presented the sample selection and its basic properties, the new ALMA observations and their data reduction, and some basic results based on the ALMA data.

The TRICEPS galaxies were selected for having \cii\ fluxes higher than 2 Jy km s$^{-1}$ and lie at the top-end of the star-forming main-sequence at $z\simeq4-5$, thus they are representative of the most massive star-forming galaxies at this cosmic epoch. The ALMA data have a mean spatial resolution of $0.1''$ ($\sim$0.8 kpc), probing the \cii\ distribution and kinematics with plenty of resolution elements (on average, 46 resolution elements per galaxy with signal-to-noise ratio higher than 3). This makes TRICEPS the largest sample of well-resolved galaxies at $z=4-5$ to date. The ALMA data are used to obtain the following results:
\begin{enumerate}
    \item In all galaxies, the gas distribution is relatively smooth with no obvious evidence for massive clumps at sub-kpc resolution. In general, \cii\ maps of massive galaxies at $z=4-5$ look remarkably similar in morphology to CO or \hi\ maps of galaxies at $z=0$ observed at similar physical resolutions.
    \item Unambiguous interaction signatures are found in three out of four galaxy pairs (BR\,1205-0725, J1551$+$0408, and SGP38326), which display tails and/or bridges in the gas and/or dust emission at large radii.
    \item In all galaxies, the cold gas forms a rotating disk, suggesting that regular rotation is ubiquitous in massive star-forming galaxies when the Universe was less than 1.5 Gyr old.
\end{enumerate}
The TRICEPS dataset will be fully exploited in future papers.

\begin{acknowledgements}
We thank the anonymous referee for constructive comments. Federico Lelli acknowledges support from the INAF mini-grant 2023 ``Galaxy dynamics across cosmic time: from statistical HI samples at z = 0 to primeval galaxies at z > 4'' and from the INAF Large Grant 2024 ``TRICEPS: luminous and dark matter in massive galaxies at z=4-5''. Cecilia Bacchini acknowledges support from the Carlsberg Foundation Fellowship Programme by Carlsbergfondet. Joe Banghal and Allison Man acknowledge the support of  the Canadian Space Agency (CSA) [23JWGO2B05] and the Natural Sciences and Engineering Research Council of Canada (NSERC) through grant reference number RGPIN- 2021-03046. Marco Castellano acknowledges support from the INAF GO Grant 2024 ``Revealing the nature of bright galaxies at cosmic dawn with deep JWST spectroscopy''. Paola Santini acknowledge support from INAF Large Grant 2024 ``UNDUST: UNveiling the Dawn of the Universe with JWST''. This paper makes use of the following ALMA data: ADS/JAO.ALMA$\#$2019.1.01587.S, $\#$2017.1.00394.S, $\#$2017.1.00127.S, $\#$2017.1.01052.S, $\#$2017.1.01471.S, $\#$2018.1.00001.S, $\#$2015.1.01345.S, $\#$2011.0.00124.S, $\#$2015.1.01564.S, $\#$2015.1.00330.S. ALMA is a partnership of ESO (representing its member states), NSF (USA) and NINS (Japan), together with NRC (Canada), MOST and ASIAA (Taiwan), and KASI (Republic of Korea), in cooperation with the Republic of Chile. The Joint ALMA Observatory is operated by ESO, AUI/NRAO and NAOJ. 
\end{acknowledgements}

\bibliographystyle{aa}
\bibliography{highz}

\begin{thebibliography}{88}
\expandafter\ifx\csname natexlab\endcsname\relax\def\natexlab#1{#1}\fi

\bibitem[{{Bacchini} {et~al.}(2019){Bacchini}, {Fraternali}, {Iorio}, \&
  {Pezzulli}}]{Bacchini2019}
{Bacchini}, C., {Fraternali}, F., {Iorio}, G., \& {Pezzulli}, G. 2019, \aap,
  622, A64

\bibitem[{{Bacchini} {et~al.}(2024){Bacchini}, {Nipoti}, {Iorio},
  {Roman-Oliveira}, {Rizzo}, {Mancera Pi{\~n}a}, {Marasco}, {Zanella}, \&
  {Lelli}}]{Bacchini2024}
{Bacchini}, C., {Nipoti}, C., {Iorio}, G., {et~al.} 2024, \aap, 687, A115

\bibitem[{{Bolatto} {et~al.}(2017){Bolatto}, {Wong}, {Utomo}, {Blitz}, {Vogel},
  {S{\'a}nchez}, {Barrera-Ballesteros}, {Cao}, {Colombo}, {Dannerbauer},
  {Garc{\'\i}a-Benito}, {Herrera-Camus}, {Husemann}, {Kalinova}, {Leroy},
  {Leung}, {Levy}, {Mast}, {Ostriker}, {Rosolowsky}, {Sandstrom}, {Teuben},
  {van de Ven}, \& {Walter}}]{Bolatto2017}
{Bolatto}, A.~D., {Wong}, T., {Utomo}, D., {et~al.} 2017, \apj, 846, 159

\bibitem[{{Bouwens} {et~al.}(2022){Bouwens}, {Smit}, {Schouws}, {Stefanon},
  {Bowler}, {Endsley}, {Gonzalez}, {Inami}, {Stark}, {Oesch}, {Hodge},
  {Aravena}, {da Cunha}, {Dayal}, {de Looze}, {Ferrara}, {Fudamoto},
  {Graziani}, {Li}, {Nanayakkara}, {Pallottini}, {Schneider}, {Sommovigo},
  {Topping}, {van der Werf}, {Algera}, {Barrufet}, {Hygate}, {Labb{\'e}},
  {Riechers}, \& {Witstok}}]{Bouwens2022}
{Bouwens}, R.~J., {Smit}, R., {Schouws}, S., {et~al.} 2022, \apj, 931, 160

\bibitem[{{Carilli} {et~al.}(2013){Carilli}, {Riechers}, {Walter}, {Maiolino},
  {Wagg}, {Lentati}, {McMahon}, \& {Wolfe}}]{Carilli2013}
{Carilli}, C.~L., {Riechers}, D., {Walter}, F., {et~al.} 2013, \apj, 763, 120

\bibitem[{{Carilli} \& {Walter}(2013)}]{CarilliWalter2013}
{Carilli}, C.~L. \& {Walter}, F. 2013, \araa, 51, 105

\bibitem[{{Carniani} {et~al.}(2013){Carniani}, {Marconi}, {Biggs}, {Cresci},
  {Cupani}, {D'Odorico}, {Humphreys}, {Maiolino}, {Mannucci}, {Molaro},
  {Nagao}, {Testi}, \& {Zwaan}}]{Carniani2013}
{Carniani}, S., {Marconi}, A., {Biggs}, A., {et~al.} 2013, \aap, 559, A29

\bibitem[{{Crawford} {et~al.}(1985){Crawford}, {Genzel}, {Townes}, \&
  {Watson}}]{Crawford1985}
{Crawford}, M.~K., {Genzel}, R., {Townes}, C.~H., \& {Watson}, D.~M. 1985,
  \apj, 291, 755

\bibitem[{{Davis} {et~al.}(2013){Davis}, {Alatalo}, {Bureau}, {Cappellari},
  {Scott}, {Young}, {Blitz}, {Crocker}, {Bayet}, {Bois}, {Bournaud}, {Davies},
  {de Zeeuw}, {Duc}, {Emsellem}, {Khochfar}, {Krajnovi{\'c}}, {Kuntschner},
  {Lablanche}, {McDermid}, {Morganti}, {Naab}, {Oosterloo}, {Sarzi}, {Serra},
  \& {Weijmans}}]{Davis2013}
{Davis}, T.~A., {Alatalo}, K., {Bureau}, M., {et~al.} 2013, \mnras, 429, 534

\bibitem[{{Davis} {et~al.}(2016){Davis}, {Greene}, {Ma}, {Pandya}, {Blakeslee},
  {McConnell}, \& {Thomas}}]{Davis2016}
{Davis}, T.~A., {Greene}, J., {Ma}, C.-P., {et~al.} 2016, \mnras, 455, 214

\bibitem[{{De Breuck} {et~al.}(2014){De Breuck}, {Williams}, {Swinbank},
  {Caselli}, {Coppin}, {Davis}, {Maiolino}, {Nagao}, {Smail}, {Walter},
  {Wei{\ss}}, \& {Zwaan}}]{DeBreuck2014}
{De Breuck}, C., {Williams}, R.~J., {Swinbank}, M., {et~al.} 2014, \aap, 565,
  A59

\bibitem[{{De Looze} {et~al.}(2014){De Looze}, {Cormier}, {Lebouteiller},
  {Madden}, {Baes}, {Bendo}, {Boquien}, {Boselli}, {Clements}, {Cortese},
  {Cooray}, {Galametz}, {Galliano}, {Graci{\'a}-Carpio}, {Isaak}, {Karczewski},
  {Parkin}, {Pellegrini}, {R{\'e}my-Ruyer}, {Spinoglio}, {Smith}, \&
  {Sturm}}]{DeLooze2014}
{De Looze}, I., {Cormier}, D., {Lebouteiller}, V., {et~al.} 2014, \aap, 568,
  A62

\bibitem[{{Di Teodoro} \& {Fraternali}(2015)}]{DiTeodoro2015}
{Di Teodoro}, E.~M. \& {Fraternali}, F. 2015, \mnras, 451, 3021

\bibitem[{{Di Teodoro} {et~al.}(2016){Di Teodoro}, {Fraternali}, \&
  {Miller}}]{DiTeodoro2016}
{Di Teodoro}, E.~M., {Fraternali}, F., \& {Miller}, S.~H. 2016, \aap, 594, A77

\bibitem[{{Di Teodoro} {et~al.}(2023){Di Teodoro}, {Posti}, {Fall}, {Ogle},
  {Jarrett}, {Appleton}, {Cluver}, {Haynes}, \& {Lisenfeld}}]{DiTeodoro2023}
{Di Teodoro}, E.~M., {Posti}, L., {Fall}, S.~M., {et~al.} 2023, \mnras, 518,
  6340

\bibitem[{{Di Teodoro} {et~al.}(2021){Di Teodoro}, {Posti}, {Ogle}, {Fall}, \&
  {Jarrett}}]{DiTeodoro2021}
{Di Teodoro}, E.~M., {Posti}, L., {Ogle}, P.~M., {Fall}, S.~M., \& {Jarrett},
  T. 2021, \mnras, 507, 5820

\bibitem[{{F{\"o}rster Schreiber} {et~al.}(2009){F{\"o}rster Schreiber},
  {Genzel}, {Bouch{\'e}}, {Cresci}, {Davies}, {Buschkamp}, {Shapiro},
  {Tacconi}, {Hicks}, {Genel}, {Shapley}, {Erb}, {Steidel}, {Lutz},
  {Eisenhauer}, {Gillessen}, {Sternberg}, {Renzini}, {Cimatti}, {Daddi},
  {Kurk}, {Lilly}, {Kong}, {Lehnert}, {Nesvadba}, {Verma}, {McCracken},
  {Arimoto}, {Mignoli}, \& {Onodera}}]{Forster2009}
{F{\"o}rster Schreiber}, N.~M., {Genzel}, R., {Bouch{\'e}}, N., {et~al.} 2009,
  \apj, 706, 1364

\bibitem[{{F{\"o}rster Schreiber} {et~al.}(2011){F{\"o}rster Schreiber},
  {Shapley}, {Genzel}, {Bouch{\'e}}, {Cresci}, {Davies}, {Erb}, {Genel},
  {Lutz}, {Newman}, {Shapiro}, {Steidel}, {Sternberg}, \&
  {Tacconi}}]{ForsterSchreiber2011}
{F{\"o}rster Schreiber}, N.~M., {Shapley}, A.~E., {Genzel}, R., {et~al.} 2011,
  \apj, 739, 45

\bibitem[{{Fujimoto} {et~al.}(2020){Fujimoto}, {Silverman}, {Bethermin},
  {Ginolfi}, {Jones}, {Le F{\`e}vre}, {Dessauges-Zavadsky}, {Rujopakarn},
  {Faisst}, {Fudamoto}, {Cassata}, {Morselli}, {Maiolino}, {Schaerer}, {Capak},
  {Yan}, {Vallini}, {Toft}, {Loiacono}, {Zamorani}, {Talia}, {Narayanan},
  {Hathi}, {Lemaux}, {Boquien}, {Amorin}, {Ibar}, {Koekemoer},
  {M{\'e}ndez-Hern{\'a}ndez}, {Bardelli}, {Vergani}, {Zucca}, {Romano}, \&
  {Cimatti}}]{Fujimoto2020}
{Fujimoto}, S., {Silverman}, J.~D., {Bethermin}, M., {et~al.} 2020, \apj, 900,
  1

\bibitem[{{Gilli} {et~al.}(2011){Gilli}, {Su}, {Norman}, {Vignali}, {Comastri},
  {Tozzi}, {Rosati}, {Stiavelli}, {Brandt}, {Xue}, {Luo}, {Castellano},
  {Fontana}, {Fiore}, {Mainieri}, \& {Ptak}}]{Gilli2011}
{Gilli}, R., {Su}, J., {Norman}, C., {et~al.} 2011, \apjl, 730, L28

\bibitem[{{Ginolfi} {et~al.}(2020){Ginolfi}, {Jones}, {B{\'e}thermin},
  {Fudamoto}, {Loiacono}, {Fujimoto}, {Le F{\'e}vre}, {Faisst}, {Schaerer},
  {Cassata}, {Silverman}, {Yan}, {Capak}, {Bardelli}, {Boquien}, {Carraro},
  {Dessauges-Zavadsky}, {Giavalisco}, {Gruppioni}, {Ibar}, {Khusanova},
  {Lemaux}, {Maiolino}, {Narayanan}, {Oesch}, {Pozzi}, {Rodighiero}, {Talia},
  {Toft}, {Vallini}, {Vergani}, \& {Zamorani}}]{Ginolfi2020}
{Ginolfi}, M., {Jones}, G.~C., {B{\'e}thermin}, M., {et~al.} 2020, \aap, 633,
  A90

\bibitem[{{Gnerucci} {et~al.}(2011){Gnerucci}, {Marconi}, {Cresci}, {Maiolino},
  {Mannucci}, {Calura}, {Cimatti}, {Cocchia}, {Grazian}, {Matteucci}, {Nagao},
  {Pozzetti}, \& {Troncoso}}]{Gnerucci2011}
{Gnerucci}, A., {Marconi}, A., {Cresci}, G., {et~al.} 2011, \aap, 528, A88

\bibitem[{{Guo} {et~al.}(2012){Guo}, {Giavalisco}, {Ferguson}, {Cassata}, \&
  {Koekemoer}}]{Guo2012}
{Guo}, Y., {Giavalisco}, M., {Ferguson}, H.~C., {Cassata}, P., \& {Koekemoer},
  A.~M. 2012, \apj, 757, 120

\bibitem[{{Heintz} {et~al.}(2022){Heintz}, {Oesch}, {Aravena}, {Bouwens},
  {Dayal}, {Ferrara}, {Fudamoto}, {Graziani}, {Inami}, {Sommovigo}, {Smit},
  {Stefanon}, {Topping}, {Pallottini}, \& {van der Werf}}]{Heintz2022}
{Heintz}, K.~E., {Oesch}, P.~A., {Aravena}, M., {et~al.} 2022, \apjl, 934, L27

\bibitem[{{Heintz} {et~al.}(2021){Heintz}, {Watson}, {Oesch}, {Narayanan}, \&
  {Madden}}]{Heintz2021}
{Heintz}, K.~E., {Watson}, D., {Oesch}, P.~A., {Narayanan}, D., \& {Madden},
  S.~C. 2021, \apj, 922, 147

\bibitem[{{Herrera-Camus} {et~al.}(2015){Herrera-Camus}, {Bolatto}, {Wolfire},
  {Smith}, {Croxall}, {Kennicutt}, {Calzetti}, {Helou}, {Walter}, {Leroy},
  {Draine}, {Brandl}, {Armus}, {Sandstrom}, {Dale}, {Aniano}, {Meidt},
  {Boquien}, {Hunt}, {Galametz}, {Tabatabaei}, {Murphy}, {Appleton}, {Roussel},
  {Engelbracht}, \& {Beirao}}]{HerreraCamus2015}
{Herrera-Camus}, R., {Bolatto}, A.~D., {Wolfire}, M.~G., {et~al.} 2015, \apj,
  800, 1

\bibitem[{{Herrera-Camus} {et~al.}(2025){Herrera-Camus},
  {Gonz{\'a}lez-L{\'o}pez}, {F{\"o}rster Schreiber}, {Aravena}, {de Looze},
  {Spilker}, {Tadaki}, {Barcos-Mu{\~n}oz}, {Assef}, {Birkin}, {Bolatto},
  {Bouwens}, {Bovino}, {Bowler}, {Calistro Rivera}, {da Cunha}, {Davies},
  {Davies}, {D{\'\i}az-Santos}, {Ferrara}, {Fisher}, {Genzel}, {Hodge},
  {Ikeda}, {Killi}, {Lee}, {Li}, {Li}, {Liu}, {Lutz}, {Mitsuhashi},
  {Narayanan}, {Naab}, {Palla}, {Price}, {Posses}, {Rela{\~n}o}, {Smit},
  {Solimano}, {Sternberg}, {Tacconi}, {Telikova}, {{\"U}bler}, {van der
  Giessen}, {Veilleux}, {Villanueva}, \&
  {Baeza-Garay}}]{HerreraCamus2025A&A...699A..80H}
{Herrera-Camus}, R., {Gonz{\'a}lez-L{\'o}pez}, J., {F{\"o}rster Schreiber}, N.,
  {et~al.} 2025, \aap, 699, A80

\bibitem[{{Hibbard} \& {van Gorkom}(1996)}]{Hibbard1996}
{Hibbard}, J.~E. \& {van Gorkom}, J.~H. 1996, \aj, 111, 655

\bibitem[{{Iorio} {et~al.}(2017){Iorio}, {Fraternali}, {Nipoti}, {Di Teodoro},
  {Read}, \& {Battaglia}}]{Iorio2017}
{Iorio}, G., {Fraternali}, F., {Nipoti}, C., {et~al.} 2017, \mnras, 466, 4159

\bibitem[{{Jones} {et~al.}(2017){Jones}, {Carilli}, {Shao}, {Wang}, {Capak},
  {Pavesi}, {Riechers}, {Karim}, {Neeleman}, \& {Walter}}]{Jones2017}
{Jones}, G.~C., {Carilli}, C.~L., {Shao}, Y., {et~al.} 2017, ArXiv e-prints
  [\eprint[arXiv]{1709.04954}]

\bibitem[{{Jones} {et~al.}(2021){Jones}, {Vergani}, {Romano}, {Ginolfi},
  {Fudamoto}, {B{\'e}thermin}, {Fujimoto}, {Lemaux}, {Morselli}, {Capak},
  {Cassata}, {Faisst}, {Le F{\`e}vre}, {Schaerer}, {Silverman}, {Yan},
  {Boquien}, {Cimatti}, {Dessauges-Zavadsky}, {Ibar}, {Maiolino}, {Rizzo},
  {Talia}, \& {Zamorani}}]{Jones2021}
{Jones}, G.~C., {Vergani}, D., {Romano}, M., {et~al.} 2021, \mnras, 507, 3540

\bibitem[{{Kapala} {et~al.}(2015){Kapala}, {Sandstrom}, {Groves}, {Croxall},
  {Kreckel}, {Dalcanton}, {Leroy}, {Schinnerer}, {Walter}, \&
  {Fouesneau}}]{Kapala2015}
{Kapala}, M.~J., {Sandstrom}, K., {Groves}, B., {et~al.} 2015, \apj, 798, 24

\bibitem[{{Kennicutt}(1998)}]{Kennicutt1998}
{Kennicutt}, Robert~C., J. 1998, \apj, 498, 541

\bibitem[{{Kennicutt} \& {Evans}(2012)}]{Kennicutt2012}
{Kennicutt}, R.~C. \& {Evans}, N.~J. 2012, \araa, 50, 531

\bibitem[{{Koprowski} {et~al.}(2024){Koprowski}, {Wijesekera}, {Dunlop},
  {McLeod}, {Micha{\l}owski}, {Lisiecki}, \& {McLure}}]{Koprowski2024}
{Koprowski}, M.~P., {Wijesekera}, J.~V., {Dunlop}, J.~S., {et~al.} 2024, arXiv
  e-prints, arXiv:2403.06575

\bibitem[{{Le F{\`e}vre} {et~al.}(2020){Le F{\`e}vre}, {B{\'e}thermin},
  {Faisst}, {Jones}, {Capak}, {Cassata}, {Silverman}, {Schaerer}, {Yan},
  {Amorin}, {Bardelli}, {Boquien}, {Cimatti}, {Dessauges-Zavadsky},
  {Giavalisco}, {Hathi}, {Fudamoto}, {Fujimoto}, {Ginolfi}, {Gruppioni},
  {Hemmati}, {Ibar}, {Koekemoer}, {Khusanova}, {Lagache}, {Lemaux}, {Loiacono},
  {Maiolino}, {Mancini}, {Narayanan}, {Morselli}, {M{\'e}ndez-Hern{\`a}ndez},
  {Oesch}, {Pozzi}, {Romano}, {Riechers}, {Scoville}, {Talia}, {Tasca},
  {Thomas}, {Toft}, {Vallini}, {Vergani}, {Walter}, {Zamorani}, \&
  {Zucca}}]{LeFevre2020}
{Le F{\`e}vre}, O., {B{\'e}thermin}, M., {Faisst}, A., {et~al.} 2020, \aap,
  643, A1

\bibitem[{{Lee} {et~al.}(2025){Lee}, {F{\"o}rster Schreiber}, {Herrera-Camus},
  {Liu}, {Price}, {Genzel}, {Tacconi}, {Lutz}, {Davies}, {Naab}, {{\"U}bler},
  {Aravena}, {Assef}, {Barcos-Mu{\~n}oz}, {Bowler}, {Burkert}, {Chen},
  {Davies}, {De Looze}, {Diaz-Santos}, {Gonz{\'a}lez-L{\'o}pez}, {Ikeda},
  {Mitsuhashi}, {Posses}, {Rela{\~n}o Pastor}, {Renzini}, {Solimano},
  {Spilker}, {Sternberg}, {Tadaki}, {Telikova}, {Veilleux}, \&
  {Villanueva}}]{Lee2025A&A...701A.260L}
{Lee}, L.~L., {F{\"o}rster Schreiber}, N.~M., {Herrera-Camus}, R., {et~al.}
  2025, \aap, 701, A260

\bibitem[{{Lelli}(2022)}]{Lelli2022}
{Lelli}, F. 2022, Nature Astronomy, 6, 35

\bibitem[{{Lelli} {et~al.}(2021){Lelli}, {Di Teodoro}, {Fraternali}, {Man},
  {Zhang}, {De Breuck}, {Davis}, \& {Maiolino}}]{Lelli2021}
{Lelli}, F., {Di Teodoro}, E.~M., {Fraternali}, F., {et~al.} 2021, Science,
  371, 713

\bibitem[{{Lelli} {et~al.}(2015){Lelli}, {Duc}, {Brinks}, {Bournaud},
  {McGaugh}, {Lisenfeld}, {Weilbacher}, {Boquien}, {Revaz}, {Braine},
  {Koribalski}, \& {Belles}}]{Lelli2015}
{Lelli}, F., {Duc}, P.-A., {Brinks}, E., {et~al.} 2015, \aap, 584, A113

\bibitem[{{Lelli} {et~al.}(2014){Lelli}, {Verheijen}, \&
  {Fraternali}}]{Lelli2014c}
{Lelli}, F., {Verheijen}, M., \& {Fraternali}, F. 2014, \mnras, 445, 1694

\bibitem[{{Lelli} {et~al.}(2012){Lelli}, {Verheijen}, {Fraternali}, \&
  {Sancisi}}]{Lelli2012a}
{Lelli}, F., {Verheijen}, M., {Fraternali}, F., \& {Sancisi}, R. 2012, \aap,
  537, A72

\bibitem[{{Lin} {et~al.}(2020){Lin}, {Ellison}, {Pan}, {Thorp}, {Su},
  {S{\'a}nchez}, {Belfiore}, {Bothwell}, {Bundy}, {Chen}, {Concas}, {Hsieh},
  {Hsieh}, {Li}, {Maiolino}, {Masters}, {Newman}, {Rowlands}, {Shi},
  {Smethurst}, {Stark}, {Xiao}, \& {Yu}}]{Lin2020}
{Lin}, L., {Ellison}, S.~L., {Pan}, H.-A., {et~al.} 2020, \apj, 903, 145

\bibitem[{{Lo} {et~al.}(1993){Lo}, {Sargent}, \& {Young}}]{Lo1993}
{Lo}, K.~Y., {Sargent}, W.~L.~W., \& {Young}, K. 1993, \aj, 106, 507

\bibitem[{{Marasco} {et~al.}(2025){Marasco}, {de Blok}, {Maccagni},
  {Fraternali}, {Oman}, {Oosterloo}, {Combes}, {McGaugh}, {Kamphuis},
  {Spekkens}, {Kleiner}, {Veronese}, {Amram}, {Chemin}, \&
  {Brinks}}]{Marasco2025}
{Marasco}, A., {de Blok}, W.~J.~G., {Maccagni}, F.~M., {et~al.} 2025, \aap,
  697, A86

\bibitem[{{McMullin} {et~al.}(2007){McMullin}, {Waters}, {Schiebel}, {Young},
  \& {Golap}}]{McMullin2007}
{McMullin}, J.~P., {Waters}, B., {Schiebel}, D., {Young}, W., \& {Golap}, K.
  2007, in Astronomical Society of the Pacific Conference Series, Vol. 376,
  Astronomical Data Analysis Software and Systems XVI, ed. R.~A. {Shaw},
  F.~{Hill}, \& D.~J. {Bell}, 127

\bibitem[{{McNichols} {et~al.}(2016){McNichols}, {Teich}, {Nims}, {Cannon},
  {Adams}, {Bernstein-Cooper}, {Giovanelli}, {Haynes}, {J{\'o}zsa}, {McQuinn},
  {Salzer}, {Skillman}, {Warren}, {Dolphin}, {Elson}, {Haurberg}, {Ott},
  {Saintonge}, {Cave}, {Hagen}, {Huang}, {Janowiecki}, {Marshall}, {Thomann},
  \& {Van Sistine}}]{McNichols2016}
{McNichols}, A.~T., {Teich}, Y.~G., {Nims}, E., {et~al.} 2016, \apj, 832, 89

\bibitem[{{Neeleman} {et~al.}(2020){Neeleman}, {Prochaska}, {Kanekar}, \&
  {Rafelski}}]{Neeleman2020}
{Neeleman}, M., {Prochaska}, J.~X., {Kanekar}, N., \& {Rafelski}, M. 2020,
  \nat, 581, 269

\bibitem[{{Noordermeer} {et~al.}(2005){Noordermeer}, {van der Hulst},
  {Sancisi}, {Swaters}, \& {van Albada}}]{Noordermeer2005}
{Noordermeer}, E., {van der Hulst}, J.~M., {Sancisi}, R., {Swaters}, R.~A., \&
  {van Albada}, T.~S. 2005, \aap, 442, 137

\bibitem[{{Ostriker} \& {Kim}(2022)}]{Ostriker2022}
{Ostriker}, E.~C. \& {Kim}, C.-G. 2022, \apj, 936, 137

\bibitem[{{Perna} {et~al.}(2022){Perna}, {Arribas}, {Colina}, {Pereira
  Santaella}, {Lamperti}, {Di Teodoro}, {{\"U}bler}, {Costantin}, {Maiolino},
  {Cresci}, {Bellocchi}, {Catal{\'a}n-Torrecilla}, {Cazzoli}, \& {Piqueras
  L{\'o}pez}}]{Perna2022}
{Perna}, M., {Arribas}, S., {Colina}, L., {et~al.} 2022, \aap, 662, A94

\bibitem[{{Pineda} {et~al.}(2013){Pineda}, {Langer}, {Velusamy}, \&
  {Goldsmith}}]{Pineda2013}
{Pineda}, J.~L., {Langer}, W.~D., {Velusamy}, T., \& {Goldsmith}, P.~F. 2013,
  \aap, 554, A103

\bibitem[{{Planck Collaboration} {et~al.}(2020){Planck Collaboration},
  {Aghanim}, {Akrami}, {Ashdown}, {Aumont}, {Baccigalupi}, {Ballardini},
  {Banday}, {Barreiro}, {Bartolo}, {Basak}, {Battye}, {Benabed}, {Bernard},
  {Bersanelli}, {Bielewicz}, {Bock}, {Bond}, {Borrill}, {Bouchet}, {Boulanger},
  {Bucher}, {Burigana}, {Butler}, {Calabrese}, {Cardoso}, {Carron},
  {Challinor}, {Chiang}, {Chluba}, {Colombo}, {Combet}, {Contreras}, {Crill},
  {Cuttaia}, {de Bernardis}, {de Zotti}, {Delabrouille}, {Delouis}, {Di
  Valentino}, {Diego}, {Dor{\'e}}, {Douspis}, {Ducout}, {Dupac}, {Dusini},
  {Efstathiou}, {Elsner}, {En{\ss}lin}, {Eriksen}, {Fantaye}, {Farhang},
  {Fergusson}, {Fernandez-Cobos}, {Finelli}, {Forastieri}, {Frailis},
  {Fraisse}, {Franceschi}, {Frolov}, {Galeotta}, {Galli}, {Ganga},
  {G{\'e}nova-Santos}, {Gerbino}, {Ghosh}, {Gonz{\'a}lez-Nuevo}, {G{\'o}rski},
  {Gratton}, {Gruppuso}, {Gudmundsson}, {Hamann}, {Handley}, {Hansen},
  {Herranz}, {Hildebrandt}, {Hivon}, {Huang}, {Jaffe}, {Jones}, {Karakci},
  {Keih{\"a}nen}, {Keskitalo}, {Kiiveri}, {Kim}, {Kisner}, {Knox},
  {Krachmalnicoff}, {Kunz}, {Kurki-Suonio}, {Lagache}, {Lamarre}, {Lasenby},
  {Lattanzi}, {Lawrence}, {Le Jeune}, {Lemos}, {Lesgourgues}, {Levrier},
  {Lewis}, {Liguori}, {Lilje}, {Lilley}, {Lindholm}, {L{\'o}pez-Caniego},
  {Lubin}, {Ma}, {Mac{\'\i}as-P{\'e}rez}, {Maggio}, {Maino}, {Mandolesi},
  {Mangilli}, {Marcos-Caballero}, {Maris}, {Martin}, {Martinelli},
  {Mart{\'\i}nez-Gonz{\'a}lez}, {Matarrese}, {Mauri}, {McEwen}, {Meinhold},
  {Melchiorri}, {Mennella}, {Migliaccio}, {Millea}, {Mitra},
  {Miville-Desch{\^e}nes}, {Molinari}, {Montier}, {Morgante}, {Moss}, {Natoli},
  {N{\o}rgaard-Nielsen}, {Pagano}, {Paoletti}, {Partridge}, {Patanchon},
  {Peiris}, {Perrotta}, {Pettorino}, {Piacentini}, {Polastri}, {Polenta},
  {Puget}, {Rachen}, {Reinecke}, {Remazeilles}, {Renzi}, {Rocha}, {Rosset},
  {Roudier}, {Rubi{\~n}o-Mart{\'\i}n}, {Ruiz-Granados}, {Salvati}, {Sandri},
  {Savelainen}, {Scott}, {Shellard}, {Sirignano}, {Sirri}, {Spencer},
  {Sunyaev}, {Suur-Uski}, {Tauber}, {Tavagnacco}, {Tenti}, {Toffolatti},
  {Tomasi}, {Trombetti}, {Valenziano}, {Valiviita}, {Van Tent}, {Vibert},
  {Vielva}, {Villa}, {Vittorio}, {Wandelt}, {Wehus}, {White}, {White},
  {Zacchei}, \& {Zonca}}]{Planck2018}
{Planck Collaboration}, {Aghanim}, N., {Akrami}, Y., {et~al.} 2020, \aap, 641,
  A6

\bibitem[{{Reuter} {et~al.}(2020){Reuter}, {Vieira}, {Spilker}, {Weiss},
  {Aravena}, {Archipley}, {B{\'e}thermin}, {Chapman}, {De Breuck}, {Dong},
  {Everett}, {Fu}, {Greve}, {Hayward}, {Hill}, {Hezaveh}, {Jarugula}, {Litke},
  {Malkan}, {Marrone}, {Narayanan}, {Phadke}, {Stark}, \&
  {Strandet}}]{Reuter2020ApJ...902...78R}
{Reuter}, C., {Vieira}, J.~D., {Spilker}, J.~S., {et~al.} 2020, \apj, 902, 78

\bibitem[{{Ribeiro} {et~al.}(2017){Ribeiro}, {Le F{\`e}vre}, {Cassata},
  {Garilli}, {Lemaux}, {Maccagni}, {Schaerer}, {Tasca}, {Zamorani}, {Zucca},
  {Amor{\'\i}n}, {Bardelli}, {Hathi}, {Koekemoer}, \& {Pforr}}]{Ribeiro2017}
{Ribeiro}, B., {Le F{\`e}vre}, O., {Cassata}, P., {et~al.} 2017, \aap, 608, A16

\bibitem[{{Riechers} {et~al.}(2013){Riechers}, {Bradford}, {Clements},
  {Dowell}, {P{\'e}rez-Fournon}, {Ivison}, {Bridge}, {Conley}, {Fu}, {Vieira},
  {Wardlow}, {Calanog}, {Cooray}, {Hurley}, {Neri}, {Kamenetzky}, {Aguirre},
  {Altieri}, {Arumugam}, {Benford}, {B{\'e}thermin}, {Bock}, {Burgarella},
  {Cabrera-Lavers}, {Chapman}, {Cox}, {Dunlop}, {Earle}, {Farrah}, {Ferrero},
  {Franceschini}, {Gavazzi}, {Glenn}, {Solares}, {Gurwell}, {Halpern},
  {Hatziminaoglou}, {Hyde}, {Ibar}, {Kov{\'a}cs}, {Krips}, {Lupu}, {Maloney},
  {Martinez-Navajas}, {Matsuhara}, {Murphy}, {Naylor}, {Nguyen}, {Oliver},
  {Omont}, {Page}, {Petitpas}, {Rangwala}, {Roseboom}, {Scott}, {Smith},
  {Staguhn}, {Streblyanska}, {Thomson}, {Valtchanov}, {Viero}, {Wang},
  {Zemcov}, \& {Zmuidzinas}}]{Riechers2013}
{Riechers}, D.~A., {Bradford}, C.~M., {Clements}, D.~L., {et~al.} 2013, \nat,
  496, 329

\bibitem[{{Rizzo} {et~al.}(2022){Rizzo}, {Kohandel}, {Pallottini}, {Zanella},
  {Ferrara}, {Vallini}, \& {Toft}}]{Rizzo2022}
{Rizzo}, F., {Kohandel}, M., {Pallottini}, A., {et~al.} 2022, \aap, 667, A5

\bibitem[{{Rizzo} {et~al.}(2018){Rizzo}, {Vegetti}, {Fraternali}, \& {Di
  Teodoro}}]{Rizzo2018}
{Rizzo}, F., {Vegetti}, S., {Fraternali}, F., \& {Di Teodoro}, E. 2018, \mnras,
  481, 5606

\bibitem[{{Rizzo} {et~al.}(2021){Rizzo}, {Vegetti}, {Fraternali}, {Stacey}, \&
  {Powell}}]{Rizzo2021}
{Rizzo}, F., {Vegetti}, S., {Fraternali}, F., {Stacey}, H.~R., \& {Powell}, D.
  2021, \mnras, 507, 3952

\bibitem[{{Rizzo} {et~al.}(2020){Rizzo}, {Vegetti}, {Powell}, {Fraternali},
  {McKean}, {Stacey}, \& {White}}]{Rizzo2020}
{Rizzo}, F., {Vegetti}, S., {Powell}, D., {et~al.} 2020, \nat, 584, 201

\bibitem[{{Roman-Oliveira} {et~al.}(2023){Roman-Oliveira}, {Fraternali}, \&
  {Rizzo}}]{RomanOliveira2023}
{Roman-Oliveira}, F., {Fraternali}, F., \& {Rizzo}, F. 2023, \mnras, 521, 1045

\bibitem[{{Santini} {et~al.}(2017){Santini}, {Fontana}, {Castellano}, {Di
  Criscienzo}, {Merlin}, {Amorin}, {Cullen}, {Daddi}, {Dickinson}, {Dunlop},
  {Grazian}, {Lamastra}, {McLure}, {Micha{\l}owski}, {Pentericci}, \&
  {Shu}}]{Santini2017}
{Santini}, P., {Fontana}, A., {Castellano}, M., {et~al.} 2017, \apj, 847, 76

\bibitem[{{Sargent} \& {Lo}(1986)}]{Sargent1986}
{Sargent}, W.~L.~W. \& {Lo}, K.~Y. 1986, in Star-forming Dwarf Galaxies and
  Related Objects, ed. D.~{Kunth}, T.~X. {Thuan}, J.~{Tran Thanh Van},
  J.~{Lequeux}, \& J.~{Audouze}, 253--262

\bibitem[{{Schreiber} {et~al.}(2015){Schreiber}, {Pannella}, {Elbaz},
  {B{\'e}thermin}, {Inami}, {Dickinson}, {Magnelli}, {Wang}, {Aussel}, {Daddi},
  {Juneau}, {Shu}, {Sargent}, {Buat}, {Faber}, {Ferguson}, {Giavalisco},
  {Koekemoer}, {Magdis}, {Morrison}, {Papovich}, {Santini}, \&
  {Scott}}]{Schreiber2015}
{Schreiber}, C., {Pannella}, M., {Elbaz}, D., {et~al.} 2015, \aap, 575, A74

\bibitem[{{Scoville} {et~al.}(2017){Scoville}, {Murchikova}, {Walter},
  {Vlahakis}, {Koda}, {Vanden Bout}, {Barnes}, {Hernquist}, {Sheth}, {Yun},
  {Sanders}, {Armus}, {Cox}, {Thompson}, {Robertson}, {Zschaechner}, {Tacconi},
  {Torrey}, {Hayward}, {Genzel}, {Hopkins}, {van der Werf}, \&
  {Decarli}}]{Scoville2017}
{Scoville}, N., {Murchikova}, L., {Walter}, F., {et~al.} 2017, \apj, 836, 66

\bibitem[{{Serra} {et~al.}(2012){Serra}, {Oosterloo}, {Morganti}, {Alatalo},
  {Blitz}, {Bois}, {Bournaud}, {Bureau}, {Cappellari}, {Crocker}, {Davies},
  {Davis}, {de Zeeuw}, {Duc}, {Emsellem}, {Khochfar}, {Krajnovi{\'c}},
  {Kuntschner}, {Lablanche}, {McDermid}, {Naab}, {Sarzi}, {Scott}, {Trager},
  {Weijmans}, \& {Young}}]{Serra2012}
{Serra}, P., {Oosterloo}, T., {Morganti}, R., {et~al.} 2012, \mnras, 422, 1835

\bibitem[{{Shelest} \& {Lelli}(2020)}]{Shelest2020}
{Shelest}, A. \& {Lelli}, F. 2020, \aap, 641, A31

\bibitem[{{Solimano} {et~al.}(2024{\natexlab{a}}){Solimano},
  {Gonz{\'a}lez-L{\'o}pez}, {Aravena}, {Alcalde Pampliega}, {Assef},
  {B{\'e}thermin}, {Boquien}, {Bovino}, {Casey}, {Cassata}, {da Cunha},
  {Davies}, {De Looze}, {Ding}, {D{\'\i}az-Santos}, {Faisst}, {Ferrara},
  {Fisher}, {F{\"o}rster-Schreiber}, {Fujimoto}, {Ginolfi}, {Gruppioni},
  {Guaita}, {Hathi}, {Herrera-Camus}, {Ibar}, {Inami}, {Jones}, {Koekemoer},
  {Lee}, {Li}, {Liu}, {Liu}, {Molina}, {Ogle}, {Posses}, {Pozzi}, {Rela{\~n}o},
  {Riechers}, {Romano}, {Spilker}, {Sulzenauer}, {Telikova}, {Vallini},
  {Vasan}, {Veilleux}, {Vergani}, {Villanueva}, {Wang}, {Yan}, \&
  {Zamorani}}]{Solimano2024a}
{Solimano}, M., {Gonz{\'a}lez-L{\'o}pez}, J., {Aravena}, M., {et~al.}
  2024{\natexlab{a}}, arXiv e-prints, arXiv:2407.13020

\bibitem[{{Solimano} {et~al.}(2024{\natexlab{b}}){Solimano},
  {Gonz{\'a}lez-L{\'o}pez}, {Aravena}, {Herrera-Camus}, {De Looze},
  {F{\"o}rster Schreiber}, {Spilker}, {Tadaki}, {Assef}, {Barcos-Mu{\~n}oz},
  {Davies}, {D{\'\i}az-Santos}, {Ferrara}, {Fisher}, {Guaita}, {Ikeda},
  {Johnston}, {Lutz}, {Mitsuhashi}, {Moya-Sierralta}, {Rela{\~n}o}, {Naab},
  {Posses}, {Telikova}, {{\"U}bler}, {van der Giessen}, {Veilleux}, \&
  {Villanueva}}]{Solimano2024b}
{Solimano}, M., {Gonz{\'a}lez-L{\'o}pez}, J., {Aravena}, M., {et~al.}
  2024{\natexlab{b}}, \aap, 689, A145

\bibitem[{{Stacey} {et~al.}(1991){Stacey}, {Geis}, {Genzel}, {Lugten},
  {Poglitsch}, {Sternberg}, \& {Townes}}]{Stacey1991}
{Stacey}, G.~J., {Geis}, N., {Genzel}, R., {et~al.} 1991, \apj, 373, 423

\bibitem[{{Stacey} {et~al.}(2010){Stacey}, {Hailey-Dunsheath}, {Ferkinhoff},
  {Nikola}, {Parshley}, {Benford}, {Staguhn}, \& {Fiolet}}]{Stacey2010}
{Stacey}, G.~J., {Hailey-Dunsheath}, S., {Ferkinhoff}, C., {et~al.} 2010, \apj,
  724, 957

\bibitem[{{Stott} {et~al.}(2016){Stott}, {Swinbank}, {Johnson}, {Tiley},
  {Magdis}, {Bower}, {Bunker}, {Bureau}, {Harrison}, {Jarvis}, {Sharples},
  {Smail}, {Sobral}, {Best}, \& {Cirasuolo}}]{Stott2016}
{Stott}, J.~P., {Swinbank}, A.~M., {Johnson}, H.~L., {et~al.} 2016, \mnras,
  457, 1888

\bibitem[{{Swinbank} {et~al.}(2012){Swinbank}, {Karim}, {Smail}, {Hodge},
  {Walter}, {Bertoldi}, {Biggs}, {de Breuck}, {Chapman}, {Coppin}, {Cox},
  {Danielson}, {Dannerbauer}, {Ivison}, {Greve}, {Knudsen}, {Menten},
  {Simpson}, {Schinnerer}, {Wardlow}, {Wei{\ss}}, \& {van der
  Werf}}]{Swinbank2012}
{Swinbank}, A.~M., {Karim}, A., {Smail}, I., {et~al.} 2012, \mnras, 427, 1066

\bibitem[{{Tadaki} {et~al.}(2018){Tadaki}, {Iono}, {Yun}, {Aretxaga},
  {Hatsukade}, {Hughes}, {Ikarashi}, {Izumi}, {Kawabe}, {Kohno}, {Lee},
  {Matsuda}, {Nakanishi}, {Saito}, {Tamura}, {Ueda}, {Umehata}, {Wilson},
  {Michiyama}, {Ando}, \& {Kamieneski}}]{Tadaki2018}
{Tadaki}, K., {Iono}, D., {Yun}, M.~S., {et~al.} 2018, \nat, 560, 613

\bibitem[{{Tadaki} {et~al.}(2019){Tadaki}, {Iono}, {Hatsukade}, {Kohno}, {Lee},
  {Matsuda}, {Michiyama}, {Nakanishi}, {Nagao}, {Saito}, {Tamura}, {Ueda}, \&
  {Umehata}}]{Tadaki2019}
{Tadaki}, K.-i., {Iono}, D., {Hatsukade}, B., {et~al.} 2019, \apj, 876, 1

\bibitem[{{Tarantino} {et~al.}(2021){Tarantino}, {Bolatto}, {Herrera-Camus},
  {Harris}, {Wolfire}, {Buchbender}, {Croxall}, {Dale}, {Groves}, {Levy},
  {Riquelme}, {Smith}, \& {Stutzki}}]{Tarantino2021}
{Tarantino}, E., {Bolatto}, A.~D., {Herrera-Camus}, R., {et~al.} 2021, \apj,
  915, 92

\bibitem[{{The CASA Team} {et~al.}(2022){The CASA Team}, {Bean}, {Bhatnagar},
  {Castro}, {Donovan Meyer}, {Emonts}, {Garcia}, {Garwood}, {Golap}, {Gonzalez
  Villalba}, {Harris}, {Hayashi}, {Hoskins}, {Hsieh}, {Jagannathan},
  {Kawasaki}, {Keimpema}, {Kettenis}, {Lopez}, {Marvil}, {Masters},
  {McNichols}, {Mehringer}, {Miel}, {Moellenbrock}, {Montesino}, {Nakazato},
  {Ott}, {Petry}, {Pokorny}, {Raba}, {Rau}, {Schiebel}, {Schweighart},
  {Sekhar}, {Shimada}, {Small}, {Steeb}, {Sugimoto}, {Suoranta}, {Tsutsumi},
  {van Bemmel}, {Verkouter}, {Wells}, {Xiong}, {Szomoru}, {Griffith},
  {Glendenning}, \& {Kern}}]{CASA2022}
{The CASA Team}, {Bean}, B., {Bhatnagar}, S., {et~al.} 2022, arXiv e-prints,
  arXiv:2210.02276

\bibitem[{{Toomre}(1964)}]{Toomre1964}
{Toomre}, A. 1964, \apj, 139, 1217

\bibitem[{{Trakhtenbrot} {et~al.}(2017){Trakhtenbrot}, {Lira}, {Netzer},
  {Cicone}, {Maiolino}, \& {Shemmer}}]{Trakhtenbrot2017}
{Trakhtenbrot}, B., {Lira}, P., {Netzer}, H., {et~al.} 2017, \apj, 836, 8

\bibitem[{{Tsukui} \& {Iguchi}(2021)}]{Tsuki2021}
{Tsukui}, T. \& {Iguchi}, S. 2021, Science, 372, 1201

\bibitem[{{Ueda} {et~al.}(2014){Ueda}, {Iono}, {Yun}, {Crocker}, {Narayanan},
  {Komugi}, {Espada}, {Hatsukade}, {Kaneko}, {Matsuda}, {Tamura}, {Wilner},
  {Kawabe}, \& {Pan}}]{Ueda2014}
{Ueda}, J., {Iono}, D., {Yun}, M.~S., {et~al.} 2014, \apjs, 214, 1

\bibitem[{{Verheijen} \& {Sancisi}(2001)}]{Verheijen2001}
{Verheijen}, M.~A.~W. \& {Sancisi}, R. 2001, \aap, 370, 765

\bibitem[{{Wagg} {et~al.}(2010){Wagg}, {Carilli}, {Wilner}, {Cox}, {De Breuck},
  {Menten}, {Riechers}, \& {Walter}}]{Wagg2010}
{Wagg}, J., {Carilli}, C.~L., {Wilner}, D.~J., {et~al.} 2010, \aap, 519, L1

\bibitem[{{Wagg} {et~al.}(2012){Wagg}, {Wiklind}, {Carilli}, {Espada}, {Peck},
  {Riechers}, {Walter}, {Wootten}, {Aravena}, {Barkats}, {Cortes}, {Hills},
  {Hodge}, {Impellizzeri}, {Iono}, {Leroy}, {Mart{\'\i}n}, {Rawlings},
  {Maiolino}, {McMahon}, {Scott}, {Villard}, \& {Vlahakis}}]{Wagg2012}
{Wagg}, J., {Wiklind}, T., {Carilli}, C.~L., {et~al.} 2012, \apjl, 752, L30

\bibitem[{{Walter} {et~al.}(2016){Walter}, {Decarli}, {Aravena}, {Carilli},
  {Bouwens}, {da Cunha}, {Daddi}, {Ivison}, {Riechers}, {Smail}, {Swinbank},
  {Weiss}, {Anguita}, {Assef}, {Bacon}, {Bauer}, {Bell}, {Bertoldi}, {Chapman},
  {Colina}, {Cortes}, {Cox}, {Dickinson}, {Elbaz}, {G{\'o}nzalez-L{\'o}pez},
  {Ibar}, {Inami}, {Infante}, {Hodge}, {Karim}, {Le Fevre}, {Magnelli}, {Neri},
  {Oesch}, {Ota}, {Popping}, {Rix}, {Sargent}, {Sheth}, {van der Wel}, {van der
  Werf}, \& {Wagg}}]{Walter2016}
{Walter}, F., {Decarli}, R., {Aravena}, M., {et~al.} 2016, \apj, 833, 67

\bibitem[{{Wisnioski} {et~al.}(2019){Wisnioski}, {F{\"o}rster Schreiber},
  {Fossati}, {Mendel}, {Wilman}, {Genzel}, {Bender}, {Wuyts}, {Davies},
  {{\"U}bler}, {Bandara}, {Beifiori}, {Belli}, {Brammer}, {Chan}, {Davies},
  {Fabricius}, {Galametz}, {Lang}, {Lutz}, {Nelson}, {Momcheva}, {Price},
  {Rosario}, {Saglia}, {Seitz}, {Shimizu}, {Tacconi}, {Tadaki}, {van Dokkum},
  \& {Wuyts}}]{Wisnioski2019}
{Wisnioski}, E., {F{\"o}rster Schreiber}, N.~M., {Fossati}, M., {et~al.} 2019,
  \apj, 886, 124

\bibitem[{{Wisnioski} {et~al.}(2015){Wisnioski}, {F{\"o}rster Schreiber},
  {Wuyts}, {Wuyts}, {Bandara}, {Wilman}, {Genzel}, {Bender}, {Davies},
  {Fossati}, {Lang}, {Mendel}, {Beifiori}, {Brammer}, {Chan}, {Fabricius},
  {Fudamoto}, {Kulkarni}, {Kurk}, {Lutz}, {Nelson}, {Momcheva}, {Rosario},
  {Saglia}, {Seitz}, {Tacconi}, \& {van Dokkum}}]{Wisnioski2015}
{Wisnioski}, E., {F{\"o}rster Schreiber}, N.~M., {Wuyts}, S., {et~al.} 2015,
  \apj, 799, 209

\bibitem[{{Zanella} {et~al.}(2018){Zanella}, {Daddi}, {Magdis}, {Diaz Santos},
  {Cormier}, {Liu}, {Cibinel}, {Gobat}, {Dickinson}, {Sargent}, {Popping},
  {Madden}, {Bethermin}, {Hughes}, {Valentino}, {Rujopakarn}, {Pannella},
  {Bournaud}, {Walter}, {Wang}, {Elbaz}, \& {Coogan}}]{Zanella2018}
{Zanella}, A., {Daddi}, E., {Magdis}, G., {et~al.} 2018, \mnras, 481, 1976

\end{thebibliography}

\appendix
\begin{appendix}

\section{Galaxy Atlas}\label{sec:atlas}

The following atlas provides overview plots for each individual galaxy, providing a compact representation of the spatially resolved observations from the TRICEPS survey. The content of the galaxy atlas is as follows:
\begin{itemize}
    \item Top-left panel: Composite RGB image using the JWST NIRCam filters F444W (R), F277W (G), and F150W (B).
    \item Top-middle panel: JWST MIRI image with the F770W filter.
    \item Top-right panel: Rest-frame 160-$\mu$m continuum map; contours are at $1,2,4,8...\times3\sigma_{\rm cont}$.
    \item Bottom-left panel: \cii\ total intensity (moment zero) map; contours are at $1,2,4,8...\times3\sigma_{\rm map}$.
    \item Bottom-middle panel: \cii\ line-of-sight velocity (moment-one) map; contours are at $0,\pm60,\pm120,\pm180...\rm km\,s^{-1}$ with respect to the galaxy redshift.
    \item Bottom-right panel: \cii\ line broadening (moment two) map. Contours are at $+30, +90, +150...\rm km\,s^{-1}$.
\end{itemize}
In all panels, the circles in the bottom-left corner shows the corresponding PSFs, while the bar in the bottom-right corner corresponds to 1 kpc. The white cross corresponds to the galaxy center determined from the gas kinematics.

\begin{figure*}
\centering
\includegraphics[width=0.9\linewidth]{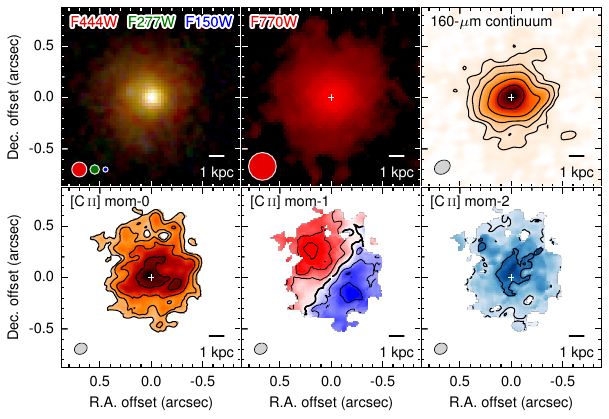}
\caption{ALESS-73.1 (SMG). See Appendix\,\ref{sec:atlas} for details.}
\label{fig: ALESS-73.1 Atlas}
\end{figure*}

\begin{figure*}
\centering
\includegraphics[width=0.9\linewidth]{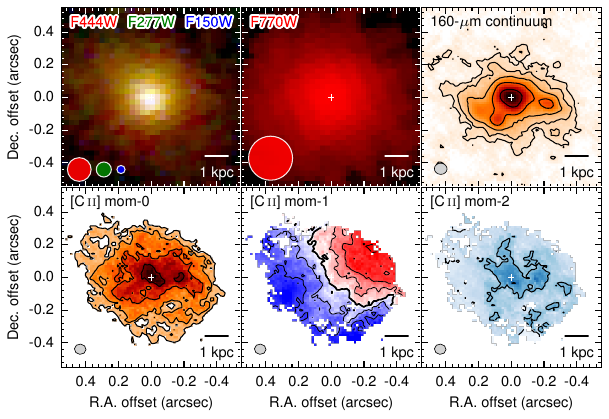}
\caption{AzTEC-1 (SMG). See Appendix\,\ref{sec:atlas} for details.}
\label{fig: AzTEC1 Atlas}
\end{figure*}

\begin{figure*}
\centering
\includegraphics[width=0.9\linewidth]{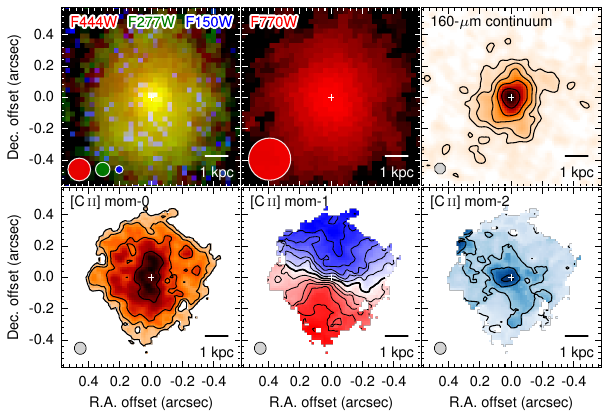}
\caption{AzTEC-159 (SMG). See Appendix\,\ref{sec:atlas} for details.}
\label{fig: AzTEC-159 Atlas}
\end{figure*}

\begin{figure*}
\centering
\includegraphics[width=0.9\linewidth]{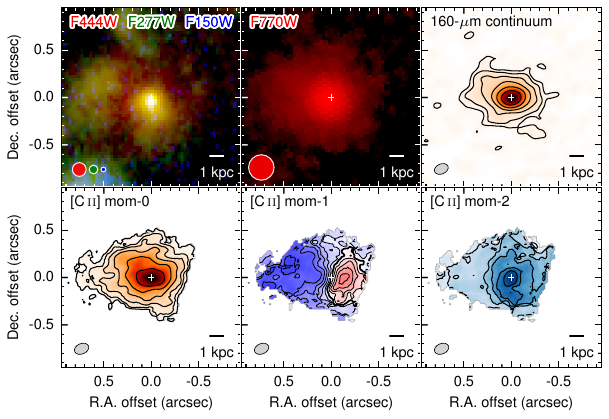}
\caption{BR1202-0725N (SMG). See Appendix\,\ref{sec:atlas} for details.}
\label{fig: BR1202-0725N Atlas}
\end{figure*}

\begin{figure*}
\centering
\includegraphics[width=0.9\linewidth]{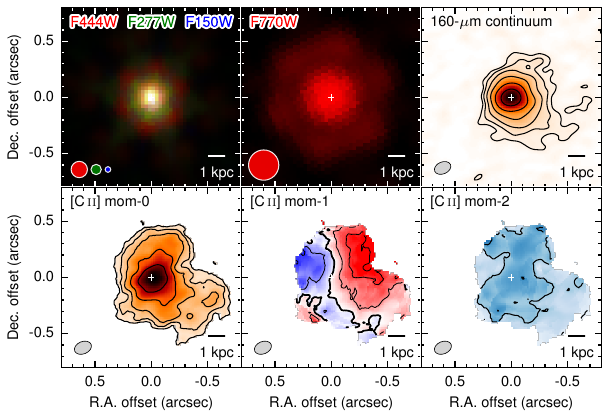}
\caption{BR1202-0725S (QSO). See Appendix\,\ref{sec:atlas} for details.}
\label{fig: BR1202-0725S Atlas}
\end{figure*}

\begin{figure*}
\centering
\includegraphics[width=0.9\linewidth]{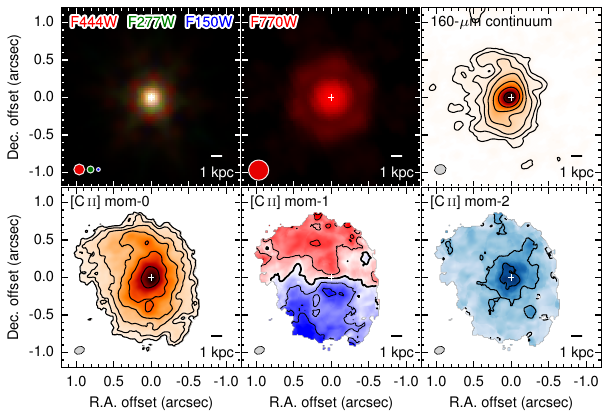}
\caption{BRI1335 (QSO). See Appendix\,\ref{sec:atlas} for details.}
\label{fig: BRI1335 Atlas}
\end{figure*}

\begin{figure*}
\centering
\includegraphics[width=0.9\linewidth]{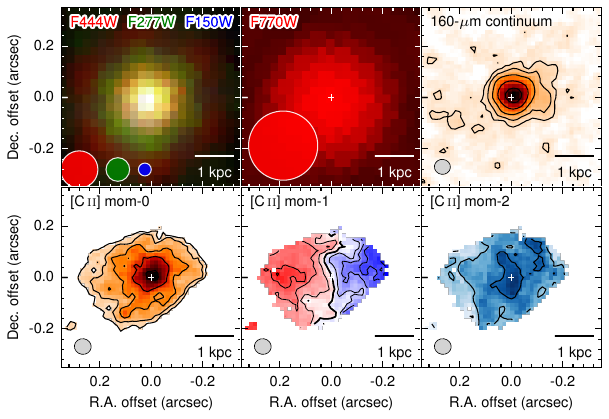}
\caption{J0331-0741 (QSO). See Appendix\,\ref{sec:atlas} for details.}
\label{fig: J0331-0741 Atlas}
\end{figure*}

\begin{figure*}
\centering
\includegraphics[width=0.9\linewidth]{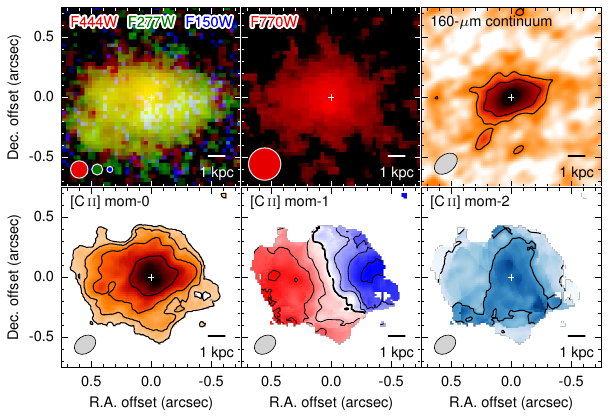}
\caption{J0817+1351 (DLA). See Appendix\,\ref{sec:atlas} for details.}
\label{fig: J0817+1351 Atlas}
\end{figure*}

\begin{figure*}
\centering
\includegraphics[width=0.9\linewidth]{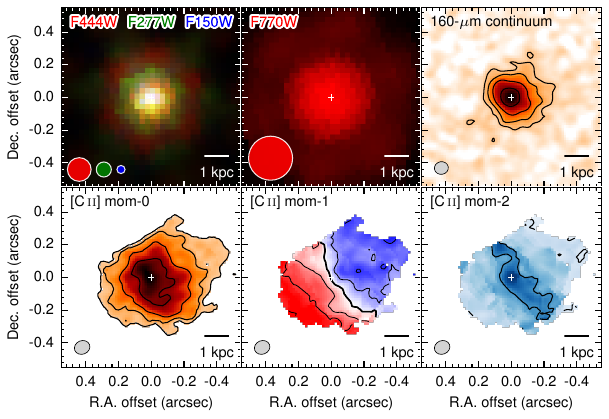}
\caption{J0923+0247N (QSO). See Appendix\,\ref{sec:atlas} for details.}
\label{fig: J0923+0247N Atlas}
\end{figure*}

\begin{figure*}
\centering
\includegraphics[width=0.9\linewidth]{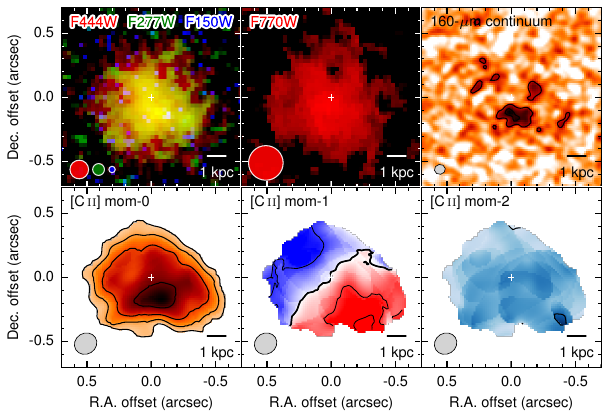}
\caption{J0923+0247S (SMG). See Appendix\,\ref{sec:atlas} for details.}
\label{fig: J0923+0247S Atlas}
\end{figure*}

\begin{figure*}
\centering
\includegraphics[width=0.9\linewidth]{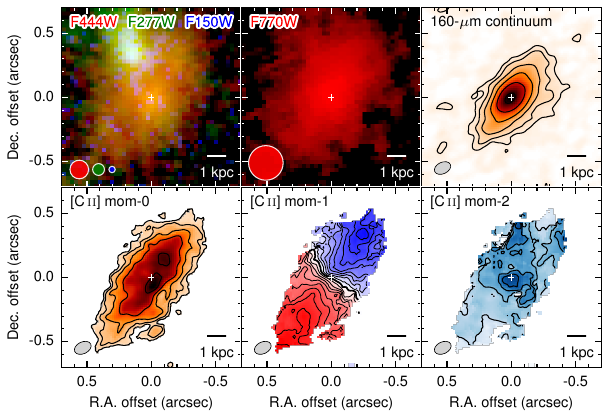}
\caption{J1000+0234 (SMG). See Appendix\,\ref{sec:atlas} for details.}
\label{fig: J1000+0234 Atlas}
\end{figure*}

\begin{figure*}
\centering
\includegraphics[width=0.9\linewidth]{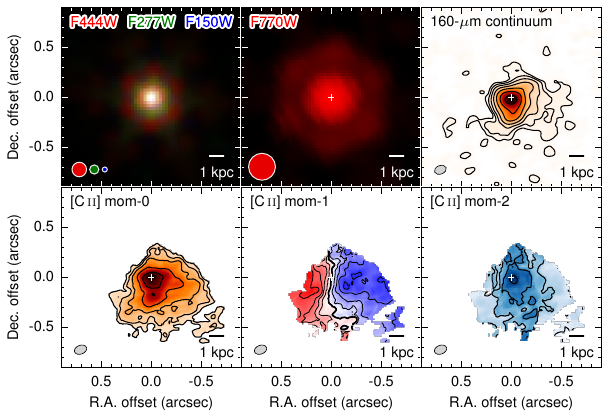}
\caption{J1341+0141 (QSO). See Appendix\,\ref{sec:atlas} for details.}
\label{fig: J1341+0141 Atlas}
\end{figure*}

\begin{figure*}
\centering
\includegraphics[width=0.9\linewidth]{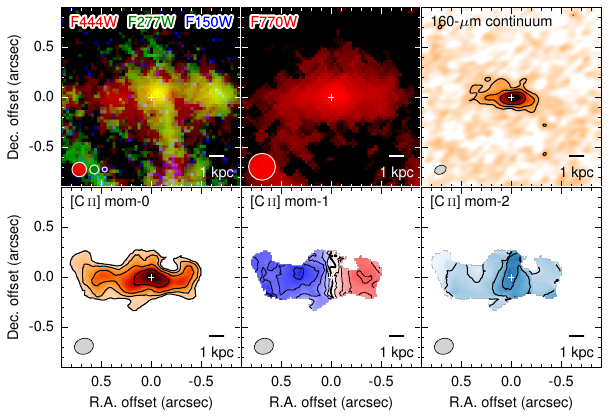}
\caption{J1511+0408N (SMG). See Appendix\,\ref{sec:atlas} for details.}
\label{fig: J1511+0408N Atlas}
\end{figure*}

\begin{figure*}
\centering
\includegraphics[width=0.9\linewidth]{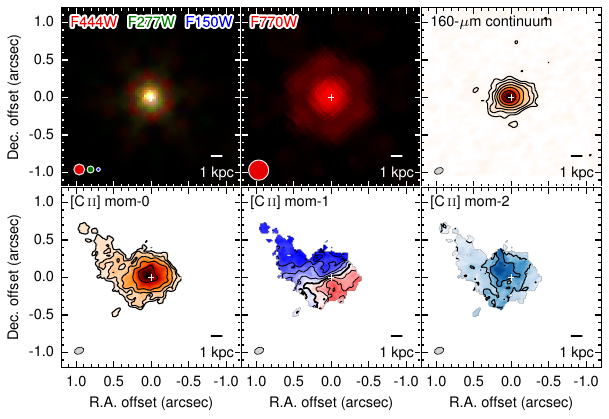}
\caption{J1511+0408S (QSO). See Appendix\,\ref{sec:atlas} for details.}
\label{fig: J1511+0408S Atlas}
\end{figure*}

\begin{figure*}
\centering
\includegraphics[width=0.9\linewidth]{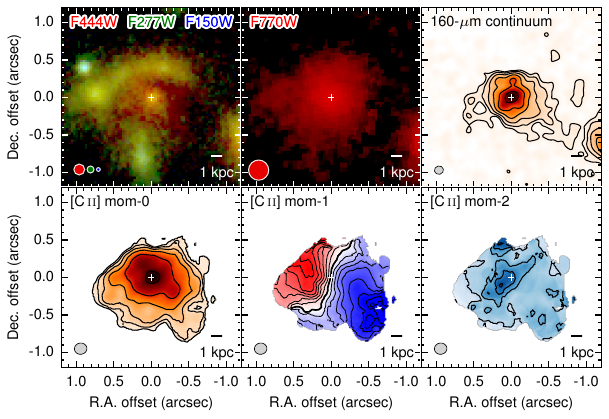}
\caption{SGP38326-1 (SMG). See Appendix\,\ref{sec:atlas} for details.}
\label{fig: SGP38326-1 Atlas}
\end{figure*}

\begin{figure*}
\centering
\includegraphics[width=0.9\linewidth]{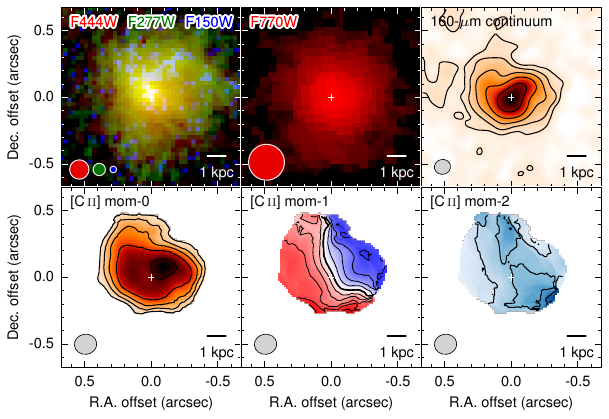}
\caption{SGP38326-2 (SMG). See Appendix\,\ref{sec:atlas} for details.}
\label{fig: SGP38326-2 Atlas}
\end{figure*}

\end{appendix}

\end{document}